\documentclass[useAMS,usenatbib]{mn2e}

\usepackage[pdf]{pstricks}
\usepackage{amsmath}
\usepackage{threeparttable}
\usepackage[]{txfonts}
\def\simless{\mathbin{\lower 3pt\hbox
{$\rlap{\raise 5pt\hbox{$\char'074$}}\mathchar"7218$}}}   
\def\simmore{\mathbin{\lower 3pt\hbox
{$\rlap{\raise 5pt\hbox{$\char'076$}}\mathchar"7218$}}}   

\newcommand{\eqb}{\begin{eqnarray}}
\newcommand{\eqe}{\end{eqnarray}}
\newcommand{\mel}{m_{\rm e}}
\newcommand{\mpr}{m_{\rm p}}

\newcommand{\gpmx}{\gamma_{\rm p, \max}}

\newcommand{\pg}{ p \pi}

\newcommand{\es}{\epsilon_{\rm s}}
\newcommand{\eo}{\epsilon_{\pi^0 \rightarrow 2\gamma}}
\newcommand{\Lgpi}{L_{\pi^0 \rightarrow 2\gamma}}
\newcommand{\Lgpeak}{L_{\gamma, \rm pk}}

\newcommand{\tpg}{t_{\rm p \pi}}

\newcommand{\tcr}{t_{\rm cr}}
\newcommand{\sth}{\sigma_{\rm T}}
\newcommand{\tgg}{\tau_{\gamma \gamma}}
\newcommand{\eblr}{\epsilon_{\rm BLR}}
\newcommand{\rblr}{R_{\rm BLR}}

\newcommand{\lblr}{\ell_{\rm BLR}}
\newcommand{\lsyn}{\ell_{\rm syn }}

\newcommand{\blow}{\beta_{1}}
\newcommand{\bhigh}{\beta_{2}}
\newcommand{\dmin}{\delta_{\min}}
\newcommand{\tvar}{t_{\rm var}}
\newcommand{\pjet}{P_{\rm jet}}
\newcommand{\pacc}{P_{\rm acc}}

\usepackage{graphicx}
\usepackage{epsfig} 
\usepackage{epstopdf}
\usepackage{color}   

\title[Constraints on the FSRQ emission region]
{Constraints of Flat Spectrum Radio Quasars in the hadronic model: the case of 3C~273}

\author[M. Petropoulou \& S. Dimitrakoudis]
{M. Petropoulou$^{1}$\thanks{E-mail: mpetropo@purdue.edu}
\thanks{Einstein Postdoctoral Fellow} \& S. Dimitrakoudis$^2$ \\
$^{1}$Department of Physics and Astronomy, Purdue University, 525 Northwestern
Avenue, West Lafayette, IN 47907, USA\\
$^2$Institute for Astronomy, Astrophysics, Space Applications \& Remote Sensing, National Observatory of Athens, 15 236 Penteli, Greece}

\begin{document}
\date{Received / Accepted}
\pagerange{\pageref{firstpage}--\pageref{lastpage}} \pubyear{2015}

\maketitle

\label{firstpage}

\begin{abstract}
We present a method of constraining the properties of the $\gamma$-ray emitting region in flat spectrum radio quasars (FSRQs) in the one-zone proton synchrotron model, where the $\gamma$-rays are produced by synchrotron radiation of relativistic protons. We show that for low enough values of the Doppler factor $\delta$, the emission from the electromagnetic (EM) cascade which is initiated by the internal absorption of high-energy photons from photohadronic interactions may exceed the observed $\sim$GeV flux. We use that effect to derive an absolute lower limit of $\delta$; first, an analytical one, in the asymptotic limit where the external radiation from the broad line region (BLR) is negligible, and then a numerical one in the more general case that includes BLR radiation. As its energy density in the emission region depends on $\delta$ and the region's distance from the galactic center, we use the EM cascade to determine a minimum distance for each value of $\delta$. We complement 
the EM cascade constraint with one derived from variability arguments and apply our method to the FSRQ 3C~273. We find that $\delta \gtrsim 18-20$ for  $B \lesssim 30$~G and $\sim$day timescale variability; the emission region is located outside the BLR, namely at $r \gtrsim 10\rblr \sim 3$~pc; 
{the  model requires at pc-scale distances stronger magnetic fields than those
inferred from core shift observations}; while the jet power exceeds by at least one order of magnitude the accretion power. 
{In short, our results disfavour the proton synchrotron model for the FSRQ 3C 273.}
\end{abstract} 
  

\section{Introduction}
\label{intro}
Blazars are a class of Active Galactic Nuclei (AGN), whose broad-band photon spectrum
is dominated by non-thermal emission. This is believed to be  
produced within a relativistic jet oriented at a small angle with respect to the line
of sight \citep{blandfordrees78, urrypadovani95}.
The spectral energy distribution (SED) of blazars is comprised {of} two broad non-thermal components: a low-energy one, that extends from the 
radio up to the  UV or X-ray frequency range, and a high-energy one that covers the X-ray and $\gamma$-ray energy bands \citep{padovanigiommi95,fossatietal98}. 

It is commonly believed that the low-energy blazar emission is the result of
electron synchrotron radiation, with the peak frequency reflecting the maximum energy at which electrons can be accelerated
\citep[e.g.][]{giommiPlanck12}. However, the origin of their high-energy emission has not been yet settled.
Among the proposed mechanisms for $\gamma$-ray production in blazars are:  synchrotron self-Compton radiation 
\citep[e.g.][]{maraschietal92, bloommarscher96, mastkirk97}, external Compton scattering \citep[e.g.][]{dermeretal92, sikoraetal94, ghisellinimadau96},
proton synchrotron radiation \citep[]{aharonian00, mueckeprotheroe01}, and photohadronic interactions \citep[e.g.][]{mannheimbiermann92,
atoyandermer01, petrodimi15}. For {Flat Spectrum Radio Quasars} (FSRQs) in particular, 
which are characterized by large values of the so-called ``Compton dominance'', 
i.e. large {ratios} of the peak high-energy luminosity to the low-energy one,
the SSC scenario is disfavoured, while the EC and proton synchrotron scenarios
remain viable \citep[e.g.][]{sikora09, chatterjee13, boettcherreimer13}.
Besides the {radiative process} responsible for
the blazar high-energy emission, {the distance of the emission region} from the super-massive black hole that lies
in the galactic center (sub-pc vs. pc scale),  remains a matter of debate \cite[e.g.][]{blazejowski00, tavecchioetal10, poutanenstern10, 
marscheretal12,sternpoutanen14}.
In the case of FSRQs, the presence of external photon fields may be used to constrain 
the location of the $\gamma$-ray emission region, at least within the leptonic EC scenario \citep[see e.g.][]{nalewajko14}.

{In} leptonic models the external radiation field
has a primary role in the formation of the SED, as it provides the seeds for inverse Compton scattering in 
the $\gamma$-ray regime. {In leptohadronic models though, the role of external photons in producing the observed
SED is not  straightforward.} 
Besides the internally produced low-energy synchrotron photons, external photons, {e.g. from the Broad Line Region (BLR),}
act as additional targets for photohadronic interactions with the accelerated protons. 
In particular, if the emission region is located within the BLR, its energy density
as measured in the respective comoving frame will  appear boosted, thus increasing
the efficiency of photopion production. It is noteworthy that FSRQs have been suggested 
as promising sites of PeV neutrino emission \citep[e.g.][]{atoyandermer01, stecker13, murase14, ANTARES2015}.
Given the recent detection of high-energy astrophysical neutrinos \citep{icecube13,aartsen14}, leptohadronic models
pose an attractive alternative to leptonic scenarios for blazar emission \citep[e.g.][]{halzenzas97}.

In this study, we present a method 
for constraining the Doppler factor and {the} location of the high-energy emission region in FSRQs within
the proton synchrotron scenario (for constraints
in the leptonic scenario of blazar emission, see e.g.~\citealt{Dondi1995, rani13, nalewajko14, zacharias15}). 
Our method is based {on the effects of the unavoidable, additional emission} produced
through  photohadronic interactions, namely via Bethe-Heitler pair production and photopion production. In addition
to the $\gamma$-ray photons produced {by} neutral pion ($\pi^0$) decay, both channels of photohadronic interactions 
lead to the injection of highly relativistic electron/positron pairs\footnote{From this point on
we refer to them commonly as `electrons'.}, that will contribute to the photon spectrum as well. 
Depending on the parameters that describe the emission region, such as its size and the magnetic field strength,
secondary pairs lose energy preferentially through synchrotron or inverse Compton processes, and their emission
signatures may appear on the SED \citep[see e.g.][]{petromast15}. 
For the case of FSRQs, their emission emerges typically at energies much {higher than} 
the peak of the high-energy component of the SED, and is also subjected to intrinsic photon-photon absorption \citep{dermer07}.
If the optical depth for intrinsic photon-photon absorption ($\tgg$) is much larger than unity, 
then an electromagnetic (EM) cascade will be initiated, transferring
energy to the GeV-TeV energy range; it may even dominate {over} the proton synchrotron emission \citep[e.g][]{mannheim91} and, in
{such a} case, the SED may no longer resemble that of a typical FSRQ.

An important quantity in our study is the photon/particle compactness, which is a 
dimensionless measure of the photon/particle energy density. It is usually expressed
as $\ell \propto L/R$, where $L$ and $R$ are, respectively, the comoving luminosity and size of the emission region.
Roughly speaking, higher compactnesses of the internally produced synchrotron photons,
of the external photons, and of the primary injected protons, result in higher photohadronic production rates and 
higher optical depths $\tgg$ \citep[see also][]{dermer07}.  
This is another manifestation of the so-called ``compactness problem'' in blazars:
the photon compactness in the $\gamma$-ray
emission region of a blazar cannot become arbitrarily high because of the initiated EM
cascades that deform its multi-wavelength emission\footnote{This has been also pointed out in 
\citep{petromast12a} for the case where no soft photons
 are {initially} present in the region (automatic photon quenching).
}. Since there are different combinations of the Doppler factor
and of the compactness of primary particles that result in the same observed flux, it follows
that the choice of low Doppler factor values favours higher production rates of secondary pairs and higher $\tgg$.
By reversing the aforementioned argument, it is evident that for a given observed $\gamma$-ray flux, size, magnetic field
strength and photon compactness, there is a minimum Doppler factor value that can be {well} defined.
Taking into account that the total (internal and external) photon compactness depends, in turn, on the location of the emission region
in the blazar jet, we can define, for each Doppler factor value,  a minimum distance as well.

This paper is structured as follows. In Sect.~\ref{analytical} we derive an analytical expression 
of the minimum Doppler factor in the limiting case where the internal photon compactness is much larger than the external one;
this sets the most stringent limit on the Doppler factor. In Sect.~\ref{numerical} we present the model and 
the algorithm for the numerical determination
of minimum distance of the emission region. We present the results of our method when applied to the FSRQ 3C~273 in Sect.~\ref{results},
and discuss other possible constraints.  We continue in Sect.~\ref{discuss} with a discussion of our results and 
conclude in Sect.~\ref{conclusions}.

\section{Analytical approach}
\label{analytical}
{The existence of a minimum Doppler factor
($\delta_{\min}$) {in the proton synchrotron model for blazar emission} can be demonstrated with analytical arguments. 
In doing so, we will also
derive an analytical expression that will reveal
its dependence on the quantities describing
the blazar emission region. Our analysis will be focused on 
the minimal  case where the internally produced synchrotron photons are the only targets for photohadronic
interactions; this is also the case of minimum photon compactness and may be realizable if the emitting 
region is located much further out of the region of external radiation (see Fig.~\ref{sketch}).

We approximate the electron synchrotron differential luminosity as a broken power-law:
\eqb
L_{\rm s}'(\epsilon') = A_1 \epsilon'^{-\blow}H[\es'-\epsilon'] + A_2 \epsilon'^{-\bhigh}H[\epsilon' - \es'],
\eqe
where $H[x]=1$ for {$x>0$} and 0 otherwise. From this point on, we use the following convention: primed and unprimed
quantities are measured in the comoving frame
of the emission region and in the observer's frame, respectively.
The normalization constants are
\eqb
A_2 & = & A_1 \es'^{\bhigh-\blow} \\
A_1 & = & \frac{(1-\blow)(\bhigh-1)}{\bhigh-\blow}\es'^{\blow-1}L'_{\rm s}.
\eqe
where $\bhigh >1$, $\blow < 1$, $L'_{\rm s} = L_{\rm s}/\delta^4$ is the total
synchrotron luminosity, and $\es' = \es (1+z)/\delta$ is the break energy of the synchrotron spectrum, which for the particular choice
of spectral indices coincides with the synchrotron peak energy. 

The differential number density of synchrotron photons is written as
\eqb
n_{\rm s}'(\epsilon') =  \tilde{A}_1 \epsilon'^{-\Gamma_1}H[\es'-\epsilon'] +\tilde{A}_2 \epsilon'^{-\Gamma_2}H[\epsilon' - \es'],
\label{ns}
\eqe
where $\Gamma_1=\blow+1$ and $\Gamma_2 = \bhigh+1$ are the low- and high-energy photon indices, respectively, and
\eqb
\tilde{A}_{1,2} = \frac{3}{4\pi c R^2}A_{1,2}.
\eqe
The optical depth for the absorption of $\gamma$-ray photons with energy $\epsilon_1'$ is
\eqb
\tgg(\epsilon'_1) \simeq \frac{R \sigma_0 (\mel c^2)^2}{\epsilon'_1}
\int_{2(\mel c^2)^2/\epsilon'_1}^{\infty}d\epsilon' \frac{n_{\rm s}(\epsilon')}{\epsilon'}
\label{tgg-general}
\eqe
where {we approximated
the photon-photon absorption cross section as\footnote{For simplifying reasons, we neglected the logarithmic dependence
.} 
$\sigma_{\gamma \gamma}(x_1 x) \approx \sigma_0 H[x_1 x -2]/(x_1 x)$
\citep{coppiblandford90}. Here, $x_1, x$ are the photon energies in units of
$\mel c^2$ and $\sigma_0 = 0.652\sth$.}
For $\es' \epsilon'_1 \gtrsim 2 (\mel c^2)^2$ the integral simplifies into
\eqb
\tgg(\epsilon_1) \approx \frac{3\sigma_0 f(\blow, \bhigh)}{8\pi c R \delta^3 (1+z)}
\frac{L_{\rm s}}{\es}\left(\frac{\es \epsilon_1 (1+z)^2}{2(\mel c^2)^2 \delta^2}\right)^{\blow},
\label{tgg}
\eqe
where 
\eqb
f= \frac{(1-\blow)(\bhigh-1)}{(\bhigh-\blow) (1+\blow)}.
\eqe

The $\gamma$-rays produced through $\pi^0$ decay are, in principal, very energetic and can easily satisfy
the threshold criterion for photon-photon absorption on the synchrotron photons with energy $\es'$. 
This can be understood as follows. In the proton synchrotron model, the high-energy component of the blazar
SED is explained as synchrotron radiation of relativistic protons.
The maximum proton Lorentz factor is related to the peak frequency ($\nu_{\gamma}$) of the
$\gamma$-ray spectrum as 
\eqb
\gpmx & = & \left(\frac{2\pi (1+z) \nu_{\gamma} \mpr c}{q_e B \delta}\right)^{1/2}
\eqe
{or,} using indicative parameter values,
\eqb
\gpmx & = &  2.4 \times 10^8 \left(B_1 \delta_1 \right)^{-1/2} \left((1+z) \nu_{\gamma, 22}\right)^{1/2},
\label{gpmx}
\eqe
where we introduced the notation $q_x \equiv q/10^x$ in cgs units. 
The typical energy of $\gamma$-ray photons produced by neutral pion decay 
is $\eo' \simeq 0.5 \kappa_{\pg} \mpr \gamma_{\rm p} c^2 $ where $\kappa_{\pg}\simeq 0.2$ is the mean proton inelasticity; {
in fact, the inelasticity increases from $\sim m_{\pi}/\mpr (\simeq 0.14)$ close to the threshold
to $\sim 0.5$ at an energy three times larger than the threshold one \citep{stecker68, begelmanrudak90}.}
Thus, protons with Lorentz factor 
$\gpmx$ result in the production of very high-energy photons:
\eqb
\label{epi0}
\eo' \simeq 2.4 \times 10^4 \ {\rm TeV} \left(B_1 \delta_1 \right)^{-1/2} \left( (1+z) \nu_{\gamma, 22}\right)^{1/2}.
\eqe
For a fiducial synchrotron peak energy $\es=0.1$~eV, which corresponds to 
\eqb
\label{es}
\es' = 0.01 \ {\rm eV} \frac{(1+z)}{\delta_1}\frac{\es}{0.1 \ {\rm eV}},
\eqe
we find that $\eo' \es' \gg 2 (\mel c^2)^2$. By substitution of eqs.~(\ref{epi0}) and (\ref{es}) in eq.~(\ref{tgg}) we find
the respective optical depth  to be
\eqb
\tgg(\eo) \simeq  \frac{4 \times 10^3}{(1+z)}
\left(\frac{0.1 \ {\rm eV}}{\es}\right)^{1/2}
\frac{L_{\rm s, 45}}{R_{16}\delta_1^3}
\left(\frac{1+z}{\delta_1}\right)^{3/4}
\left(\frac{\nu_{\gamma, 22}}{B_1} \right)^{1/4},
\eqe
where we assumed $\blow=1/2$ and $\bhigh=3/2$.
Since $\tgg(\eo) \gg 1$ is typical, the $\gamma$-ray luminosity from $\pi^0$ decay ($\Lgpi$) will be totally absorbed.
We may thus write that $\Lgpi^{\rm abs} = \left(1-e^{-\tgg}\right)\Lgpi \simeq \Lgpi$.
The absorbed photon luminosity will be re-distributed
at lower $\gamma$-ray energies through the development of an EM cascade.
This emerges as an { additional emission that should be}  below the proton synchrotron component, which
in our framework is responsible for the FSRQ high-energy emission.

We note that photons emitted by secondary, highly relativistic electrons from charged ($\pi^{+}$) pion decay or/and Bethe-Heitler pair production are also subject to photon-photon absorption, and thus they may contribute to the cascade emission. 
{The synchrotron photons emitted  by secondary pairs are less energetic than those from
$\pi^0$ decays. Thus, they are mainly attenuated by photons
with energies $\epsilon' \gtrsim \es'$. To exemplify this, let us consider the most energetic pairs 
produced by pion decays. These are produced roughly with $\gamma_{\pg} \simeq (1/4)\kappa_{\pg} \gamma_{\rm p, \max} \mpr/\mel$, and 
the respective synchrotron photon energy is written as}
\eqb
\epsilon'_{\rm s, \pg} \simeq 60\ {\rm TeV} \ \delta_1^{-1} (1+z) \nu_{\gamma, 22}.
\label{espg}
\eqe
{This lies just above the threshold for $\gamma \gamma$ absorption on photons with $\es'$. Cooling of pairs with $\gamma_{\pg}$
as well as the production of pairs from  protons with $\gamma_{\rm p} < \gpmx$  results in photon 
emission at $\epsilon' \ll \epsilon'_{\rm s, \pg}$, where the threshold condition for absorption on $\es'$ is no more satisfied.
Since $n'_{\rm s}\propto \epsilon'^{-1-\bhigh}$ for $\epsilon'>\es'$, 
the optical depth for absorption of photons with $\epsilon' \ll \epsilon'_{\rm s, \pg}$
is expected to be much less than 
\eqb
\tgg(\epsilon'_{\rm s, \pg}) \simeq  \frac{10^2}{(1+z)}
\left(\frac{0.1 \ {\rm eV}}{\es}\right)
\frac{L_{\rm s, 45}}{R_{16}\delta_1^3},
\eqe
where we used eqs.~(\ref{espg}), (\ref{tgg-general}) and $\es' \epsilon'_{\rm s, \pg} \approx 2 (\mel c^2)^2$.
Moreover, the synchrotron spectrum from pairs spans many decades in energy and
the luminosity emitted at $\epsilon'_{\rm s, \pg}$ is, therefore, only
a fraction of the total injected luminosity in pairs. This is not the case
for the $\gamma$-ray spectrum from $\pi^0$ decays, which is sharply peaked at $\eo'$.
Similar arguments apply to the synchrotron emission from Bethe-Heitler pairs, which are produced
on average with $\gamma < \gamma_{\pg}$. Thus, for the purposes of {this} analytical approach,
we can safely ignore the attenuation of photons from Bethe-Heitler and pion decay process,
and consider only the attenuation of $\gamma$-ray photons from $\pi^0$ decay. In any case,
our analytical results will be compared against those calculated numerically,
after taking into account  the additional photon emission from Bethe-Heitler and charged pion processes (see Sect.~\ref{results}).}

{Since the proton synchrotron emission alone can explain the observed
$\gamma$-ray spectrum,  the sum of the cascade and proton synchrotron emission
may exceed the observations for high enough values of $\Lgpi$}. This can be avoided if the following energetic constraint is satisfied 
\eqb
\Lgpi^{\rm abs}\lesssim \eta \Lgpeak,
\label{constrain0}
\eqe
where $\Lgpeak$ is the peak luminosity of the high-energy SED component, which is typically a good proxy of the total $\gamma$-ray luminosity.
Here, $\eta \le 1$  is a dimensionless factor to be defined later by the observations (see Sect.~\ref{results}). 
The above relation does not take into account any spectral
information about the developed EM cascade \citep[e.g.][]{mannheim93, petroarfani13}. 
It is based on the simplifying assumption that the absorbed luminosity re-emerges at the energy where
$\tgg (\epsilon_{\star})\sim 1$.  Using fiducial parameter values and solving eq.~(\ref{tgg}) for $\epsilon_{\star}$, we find 
\eqb
\epsilon_{\star} \simeq {18}\ {\rm GeV} \frac{\es}{0.1 \ {\rm eV}} \delta_1^8 L_{\rm s, 45}^{-2} R_{16}^2,
\label{estar}
\eqe
where we also assumed $\blow=1/2$ and $\bhigh=3/2$. We thus expect the EM cascade to emerge at energies higher than the peak
of the high-energy emission, which for FSRQs usually falls in the {4~MeV-40~MeV} ($10^{21}$~Hz-$10^{22}$~Hz) range \citep{fossatietal98}.
The constraint imposed by relation (\ref{constrain0}) could be relaxed
if we were to include spectral information for the cascade emission. 
{However}, this lies out of the scope of the present work.

The $\gamma$-ray luminosity from the $\pi^0$ decay may be written as
\eqb
\Lgpi^{\rm abs}\simeq \Lgpi \simeq 2\times \frac{1}{2}\tau_{\pg}L_{\rm p},
\label{constrain1}
\eqe
where $\tau_{\pg}$ is the optical depth for photopion interactions (to be defined below) and $L_{\rm p}$
is the total injected proton luminosity. The factors $2$ and $1/2$ account for the production of two photons
that share the energy of the parent neutral pion. In what follows, we will use eqs.~(\ref{constrain0}) and (\ref{constrain1}) 
for deriving and justifying the existence of a minimum Doppler factor.

Before we calculate the optical depth for {photopion ($\pg$) interactions, it is useful
to determine the 
threshold photon energy for such interactions with 
protons having} Lorentz 
factor $\gpmx$. Using eq.~(\ref{gpmx}) we find
\eqb
\epsilon'_{\rm th} \simeq 0.58 \ {\rm eV} \left(B_1 \delta_1 \right)^{1/2} \left( (1+z) \nu_{\gamma, 22}\right)^{-1/2},
\eqe
which for typical parameter values, is $\epsilon'_{\rm th} \gg \es'$ (see also eq.~(\ref{es})). 
Lower energy protons will interact with synchrotron photons of energy $\epsilon'> \epsilon'_{\rm th}$, whose
number density decreases as $\propto \epsilon'^{-1-\bhigh}$; we will not consider 
these interactions in the following.

The optical depth for $\pg$ interactions is defined as $\tau_{\pg}(\gamma_{\rm p}) \equiv t_{\rm cr}/t_{\pg}(\gamma_{\rm p})$,
where $t_{\rm cr}=R/c$ and $\tpg^{-1}$ is the $\pg$ energy loss rate given by \citep{stecker68}
\eqb
t^{-1}_{\pg}(\gamma_{\rm p}) \approx \frac{c}{2\gamma_{\rm p}^2}
\int_{\bar{\epsilon}_{\rm th}}^{\infty}d\bar{\epsilon}\bar{\epsilon}\sigma_{\pg}(\bar{\epsilon}) \kappa_{\pg}(\bar{\epsilon})
\int_{\bar{\epsilon}/2\gamma_{\rm p}}^{\infty} d\epsilon' \frac{n'(\epsilon')}{\epsilon'^2},
\eqe
where $\bar{\epsilon}_{\rm th}=145$~MeV.
We assume $\kappa_{\pg}(\bar{\epsilon})\approx 0.2$ and approximate the cross section as $\sigma_{\pg} = 
\bar{\sigma}_{\pg}H[\bar{\epsilon}-\bar{\epsilon}_{\rm th}]$, where $\bar{\sigma}_{\pg}= 1.5\times 10^{-4} \sth$
(for a more realistic description of $\sigma_{\pg}$, see \citealt{SOPHIA2000, beringeretal12}). 
For the photon spectrum
defined by eq.~(\ref{ns}), the second integral is written
as 
\eqb
\int_{\bar{\epsilon}/2\gamma_{\rm p}}^{\infty} d\epsilon' \frac{n'(\epsilon')}{\epsilon'^2} = 
\int_{\bar{\epsilon}/2\gamma_{\rm p}}^{\es'} \!\! d\epsilon' \tilde{A}_1 \epsilon'^{-3-\blow}+
\int_{\max[\bar{\epsilon}/2\gamma_{\rm p}, \es']}^{\infty}\!\!\!\!\! d\epsilon' \tilde{A}_2 \epsilon'^{-3-\bhigh}.
\eqe
For the highest energy protons, i.e. with Lorentz factor $\gpmx$,
only the second integral is non-zero, and the respective optical depth is written
as
\eqb
\tau_{\pg}(\gpmx) \simeq \frac{3\bar{\sigma}_{\pg}\kappa_{\pg}}{2\pi cR} 
\frac{L_{\rm s}}{\es}\frac{g(\bhigh, \blow)}{\delta^3 (1+z)} \left(\frac{2\gpmx \es(1+z)}{\delta \bar{\epsilon}_{\rm th}} \right)^{\bhigh}
\label{tpg2}
\eqe
where 
\eqb
g(\bhigh, \blow) = \frac{(1-\blow)(\bhigh-1)}{\bhigh(\bhigh-\blow)(2+\bhigh)}.
\eqe
{Using eq.~(\ref{gpmx}), $\blow=1/2$, $\bhigh=3/2$ and fiducial values for the other parameters, the optical depth
is written as
\eqb
\tau_{\pg}(\gpmx) \simeq 6\times 10^{-5} \frac{L_{\rm s, 45} \nu_{\gamma, 22}^{3/4}}{R_{16} B_1^{3/4}\delta_1^{21/4}(1+z)^{1/4}}
\left(\frac{\es}{0.1 \ {\rm eV}} \right)^{1/2}.
\label{tpg3}
\eqe}
Using the approximation $\Lgpeak \approx L_{\gamma}$, the definition of the proton compactness
$\ell_{\rm p}$, which is a dimensionless measure of the proton luminosity, given by 
\eqb
\ell_{\rm p} = \frac{\sth L_{\rm p}}{4\pi R \mpr c^3 \delta^4},
\eqe
and eqs.~(\ref{constrain0}), (\ref{constrain1}) and (\ref{tpg2}) 
we find
that 
\eqb
6g(\bhigh,\blow)\frac{\bar{\sigma}_{\pg}\kappa_{\pg}}{\sth}\frac{L_{\rm s}}{\eta L_{\gamma}}\left(\es (1+z)\right)^{\bhigh-1}\left(\frac{2\gpmx}
{\bar{\epsilon}_{\rm th}} \right)^{\bhigh} \ell_{\rm p}\mpr c^2\delta^{1-\bhigh} \lesssim 1.
\label{constrain2}
\eqe
At this point, we make use of our working hypothesis, namely that the $\gamma$-ray emission
is explained
by proton synchrotron radiation. 
Using standard expressions for the synchrotron luminosity emitted by a power-law proton
distribution (e.g. eq.~(6.36) in \cite{rybicki}) we may express $\ell_{\rm p}$ as
\eqb
\ell_{\rm p} =\frac{C_{\rm p} \sth L_{\gamma}B^{-(p+1)/2} \delta^{-(p+5)/2}}{4\pi R^2 f_{\rm p} \nu_{\gamma}^{(3-p)/2}}
\label{lp}
\eqe
where $f_{\rm p}=(p-2)/(p-1)$ with $p \ne 1$ being the power-law index of the injected proton distribution and 
\eqb
\label{const_p}
C_{\rm p} &=& \frac{(3-p)(p+1)\left(2\pi\right)^{(p-1)/2} \mpr^{(p+1)/2} c^{(p+3)/2} }{2 q_e^{(5+p)/2}3^{p/2}\Gamma_1(p) \Gamma_2(p) } \\
\Gamma_1 & = &  \Gamma\left( \frac{p}{4} + \frac{19}{12}\right) \\
\Gamma_2 & = &  \Gamma\left( \frac{p}{4} - \frac{1}{12}\right),
\eqe
where $\Gamma(t) \equiv \int_0^{\infty} dx x^{t-1} e^{-x}$.
The constant $C_{\rm p}^{-1}$ is a generalization of $C_2$ given by eq.~(9) in \cite{petromast12a} for $ 2<p < 3$.
Substitution of eqs.~(\ref{gpmx}) and (\ref{lp})  into  eq.~(\ref{constrain2})
results in 
\eqb
\delta \gtrsim \delta_{\min},
\eqe
with
\eqb
\dmin^{({p+3} +3\bhigh)/2} \simeq  C_1 L_{\rm s} \eta^{-1} R^{-2} \es^{\bhigh-1}\nu_{\gamma}^{(\bhigh+p-3)/2}
B^{-(\bhigh+p+1)/2}.
\label{dmin1}
\eqe
In the above,
\eqb
C_1 = \frac{6}{4\pi}g(\bhigh, \blow)\bar{\sigma}_{\pg}\kappa_{\pg}\mpr c^2 C_{\rm p} \frac{f_{\rm z}^{3\bhigh/2}}{f_{\rm z}f_{\rm p}}
\left(\frac{2}{\bar{\epsilon}_{\rm th}} \right)^{\bhigh}\left( \frac{2\pi \mpr c}{q_e}\right)^{\bhigh/2}
\eqe
and $f_{\rm z}\equiv1+z$.
%
It is noteworthy that $\dmin$ does not depend on the $\gamma$-ray luminosity, while it has a  weak dependence on most
of other model parameters:
\eqb
\label{dependence}
\delta_{\min} \propto R^{-1/w}\eta^{-{2}/{w}} L_{\rm s}^{{2}/{w}} \es^{{2(\bhigh-1)}/{w}}
\nu_{\gamma}^{({\bhigh+p-3})/{w}}B^{-({p+1+\bhigh})/{w}},
\eqe
{where $w=p+3+3\bhigh$.}
Instead, the strongest dependence comes through the magnetic field strength and the spectral index of the synchrotron spectrum above its peak: 
\begin{itemize}
\item the minimum Doppler factor decreases for stronger magnetic fields. 
  Higher values of $B$ require lower proton luminosity to explain a given
 observed $\gamma$-ray luminosity, as eq.~(\ref{lp}) demonstrates. This subsequently reduces the luminosity produced
 through photohadronic interactions and, thus, the amount of 
energy being absorbed and reprocessed (see eq.~(\ref{constrain1}));

 \item the minimum Doppler factor decreases  as the electron synchrotron spectrum above its peak becomes steeper.
 This can be easily understood, since the respective number density of synchrotron photons scales as 
 $n'_{\rm s}(\epsilon') \propto \epsilon'^{-1-\bhigh}$ and decreases for higher $\bhigh$. 
 We remind that very high energy $\gamma$-rays produced
 via photohadronic interactions  are mostly absorbed by these synchrotron photons.
\end{itemize}

{The dependence of $\dmin$ on the magnetic field strength is exemplified in Fig.~\ref{fig1}, where 
$\delta_{\min}$ is shown as a function of $B$} for two values of the emission region radius: $R=3.6\times 10^{16}$~cm (solid line) and
$3.6\times 10^{15}$~cm (dashed line). 
Other parameters used for the plot are: $\nu_{\rm s} = \es/h = 3.2\times 10^{13}$~Hz, 
$L_{\rm s}=6.3\times10^{45}$~erg/s (this corresponds
to $10^{13}$~Jy Hz for the 3C 273 distance $D_{\rm L}=755$~Mpc), $\nu_{\gamma}=10^{22}$~Hz, $\eta=0.2$, $p=2.3$, 
$\blow=0.7$ and $\bhigh=1.35$.  Our choice of the parameter values is motivated
by the SED fitting of 3C~273 (see Sect.~\ref{results}).

\begin{figure}
 \centering
\includegraphics[width=0.48\textwidth]{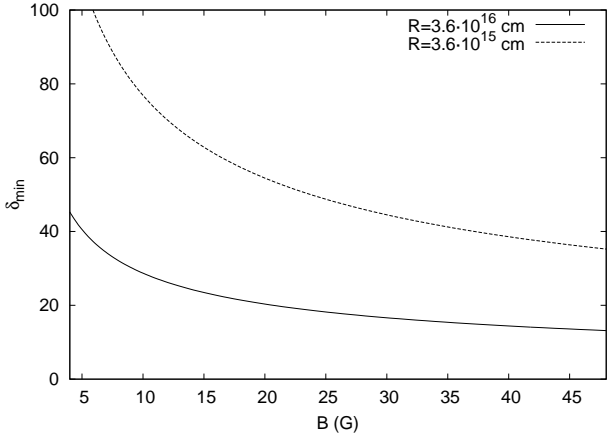}
\caption{Minimum Doppler factor $\delta_{\min}$ as a function of the magnetic field strength for
$R=3.6\times 10^{16}$~cm (solid line) and $R=3.6\times 10^{15}$~cm (dashed line). 
The region below the curves leads to modification of the proton synchrotron spectrum
due to the enhanced cascade emission, and is therefore, forbidden. Other parameters used for the plot are: $\nu_{\rm s} = \es/h = 3.2\times 10^{13}$~Hz, 
$L_{\rm s}=6.3\times10^{45}$~erg/s, $\nu_{\gamma}=10^{22}$~Hz, $\eta=0.2$, $p=2.3$, 
$\blow=0.7$ and $\bhigh=1.35$. }
\label{fig1}
\end{figure}
{
Figure \ref{fig1} demonstrates that 
the Doppler factor of the high-energy emission region in FSRQs lies above 15-20, unless
the magnetic field strength is high, i.e. $\gtrsim 30$~G and/or 
the synchrotron photon spectrum is steep, e.g. $\bhigh \gtrsim 2$. }
It is important to note that we have {arrived at} this conclusion without considering {any} additional constraints
imposed by { e.g.} the observed high-energy variability. In other words, even if the observed variability is not faster than $\sim$hour timescale, 
the cascade emission initiated by photohadronic interactions
limits the Doppler factor to values larger than 15-20. 

The minimum Doppler factor shown in Fig.~\ref{fig1} can be considered as an absolute
lower limit, since it was derived using only the internally produced radiation. If we were to include an extra low-energy 
photon component in our calculations, such as the BLR photon field, the Doppler factor would have to be larger than $\dmin$ (see Sect.~\ref{results}).

\section{Numerical approach}
\label{numerical}
In what follows we will expand upon the idea presented
in the previous section by including in our calculations 
the emission from the BLR.  
Our working framework is
analogous to that adopted in the previous section but with
two main differences: 
\begin{enumerate}
 \item the use of the numerical code described 
  in \cite{DMPR2012} allows us to make 
  no assumptions about the photohadronic emission and the initiated EM cascade.
  The steady-state proton, electron and photon distributions are
  self-consistently calculated by
  numerically solving the system of coupled
  integrodifferential equations that describes their evolution in the energy- and time-phase space.
\item the minimum Doppler factor for a particular FSRQ will be derived by fitting its multi-wavelength contemporaneous observations 
with the proton synchrotron model.
\end{enumerate}

\subsection{Model}
\label{model}
We model the blazar emission region as a spherical homogeneous blob of radius $R$
that contains a tangled magnetic field of strength B. The region 
moves with a bulk Lorentz factor $\Gamma$ at a small angle $\theta_{\rm obs}$ with respect to the observer.
The respective Doppler factor $\delta$ is defined as $\delta = \Gamma^{-1}(1-\beta \cos\theta_{\rm obs})^{-1}$.
We assume that relativistic electrons and protons with  power-law distributions
are being injected into the source at a constant rate, while they may physically escape
at an energy-independent timescale that is set equal to the crossing time $\tcr=R/c$ of the emission region.
Electrons lose energy through synchrotron radiation and inverse Compton scattering on the external photons (EC) as well
as on the internally produced synchrotron photons (SSC). 
Protons lose energy by emitting synchrotron radiation
and through the photohadronic channels of Bethe-Heitler pair production
and photopion production \citep[e.g.][]{DMPR2012}. 
The loss processes will lead to the injection of secondary electrons and photons
which are, respectively, subjected to synchrotron/inverse Compton scattering and photon-photon absorption.
We refer the reader to \cite{dpm14} for a detailed description of the physical processes.

As already mentioned in the introduction, our working hypothesis is that the low-and high-energy components
of the blazar SED are the result of primary electron and proton synchrotron radiation, respectively \citep[see e.g.][]{mueckeprotheroe01}.
The emission produced through photohadronic interactions appears at
even higher energies than the high-energy {component}, and is subjected
to photon-photon absorption. 

We assume that the spectrum of the external ultraviolet  (UV) radiation arises from 
an optically thick accretion disk, {and is} then scattered by the BLR clouds.
For simplicity, we approximate the accretion disk emission with a black-body spectrum\footnote{An optically thick, geometrically thin
disk, i.e. a Shakura-Sunyaev disk, is better described by a multi-temperature black body, whose flux
scales as $F(\epsilon)\propto \epsilon^{1/3}\exp(-\epsilon/\epsilon_{0})$. 
However, the details {of} the accretion disk spectrum do not affect our analysis.} that peaks
at the observed energy. 
For the BLR we adopt the geometry presented in \cite{nalewajko14}\footnote{Some observations may suggest, {however,} 
a planar geometry for certain FSRQs \citep[e.g.][]{sternpoutanen14}.},  and illustrated in Fig.~\ref{sketch}.
The inner radius of the BLR is defined as $\rblr$, which is related
to the accretion disk luminosity $L_{\rm ad}$ \citep{ghisellini_tavecchio08} as 
\eqb
\rblr \approx 10^{17}~{\rm cm} \ L_{\rm ad, 45}^{1/2}.
\label{rblr}
\eqe
Other sources of external photons could be the reprocessed line emission or infrared radiation from a dusty torus.
In what follows, we will not include in our calculations the radiation from the torus, 
since its luminosity and size are less well-defined than {those} for the BLR. 
We will also neglect the direct irradiation from the accretion disk \citep{dermerschlickeiser02}. This is a safe assumption as
long as the emission region lies at \citep[e.g.][]{ghisellinimadau96, sikora09} 
\eqb
r > 0.8\left(\frac{R_{\rm g} \rblr^2}{\xi_{\rm BLR}} \right)^{1/3} \simeq 0.01 \ {\rm pc} \ \xi_{\rm BLR, -1}^{-1/3} L_{\rm ad, 45}^{1/3}M_{\rm BH, 9}^{1/3},
\eqe
where $M_{\rm BH}$ is the black hole mass, which for 3C~273 is 
$M_{\rm BH, 9}\equiv M_{\rm BH}/(10^9 M_{\sun}) \simeq 0.9-2.4$ \citep{peterson04, paltani05},
and  $\xi_{\rm BLR}$ is a dimensionless factor that incorporates all the details about
the geometry and the irradiation of the BLR from the accretion disk.
A representative value is $\xi_{\rm BLR}\sim 0.1$  \citep{sikora09}, while values  $\lesssim 0.01$ are considered to be very
low (e.g. \citealt{nalewajko14}).
 
\begin{figure}
 \centering
 \includegraphics[width=0.5\textwidth]{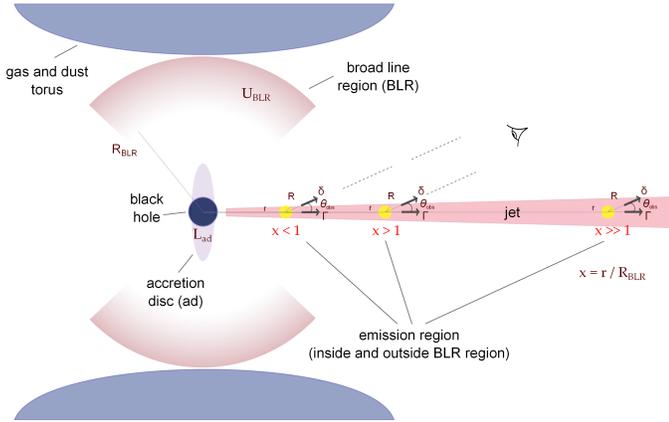}
 \caption{Sketch of the geometry of the 
 emission region and its
 surrounding sources of radiation, for an FSRQ.}
 \label{sketch}
\end{figure}
The emission region, which is depicted as a yellow blob in Fig.~\ref{sketch}, 
is located at a distance $r$ in the blazar jet. 
In this study, we treat $r$ as a free parameter, i.e. $r \lessgtr \rblr$, with the aim
of imposing a minimum value on the ratio $r/\rblr$.
As shown in Fig.~\ref{sketch}, the radius of the emission region is kept constant, with {only the} constraint
of being smaller than the {transverse size of the jet} at a distance $r$, i.e. $R \lesssim r\theta \lesssim r/\Gamma$, where
we also used $\theta\Gamma \lesssim 1$ (see \citealt{nalewajko14} and discussion, therein). 
We will return to this issue in Sect.~\ref{discuss}, where we discuss how our results
would be altered, if we allowed $R \propto r^{s}$. In principle, $R$  is related
to the observed variability timescale as $\tvar \gtrsim (1+z)R/c \delta$. As 
$\tvar$ may take a wide range of values, even for the same
source, depending on the observing period and energy band \citep[for 3C~273 see e.g.][]{kataoka02, soldi08}, we choose to use
$R$ instead of $\tvar$ as the free parameter. {We discuss the variability constraints later in Sect.~\ref{constraints}.}

Following \cite{sikora09} and \cite{nalewajko14}, we write the energy density of the BLR region as measured in the 
rest frame of the emission region as
\eqb
u'_{\rm BLR} =  \frac{0.4 \xi_{\rm BLR} \Gamma^2 L_{\rm ad}}{3\pi c \rblr^2} \lambda(x),
\label{ublr}
\eqe
where the function $\lambda$ is defined as  
\eqb
\lambda(x) & = & \frac{1}{1+x^4},
\eqe
and $x\equiv r/\rblr$. 
Similarly, the BLR photon energy as measured in the comoving frame of the emission region
is given by \citep{nalewajko14}
\eqb
\eblr' \simeq \Gamma\eblr \frac{1}{1+x^3}.
\label{eblr}
\eqe
An important quantity in our analysis is the so-called photon compactness\footnote{Similarly,  we have used in Sect.~\ref{analytical} the
term proton compactness (see eq.~(\ref{lp})).}, which
is defined
as $\ell_{\gamma}\equiv u'_{\gamma}  \sth R / \mel c^2$, where $u'_{\gamma}$ is the comoving energy
density of an {arbitrary} photon field. Using eq.~(\ref{ublr})
we may write the BLR photon compactness as
\eqb
\lblr = \ell_0\Gamma^2\lambda(x),
\label{lblr}
\eqe
where 
\eqb
\ell_ 0 =  \frac{0.4 \xi_{\rm BLR}\sth R L_{\rm ad}}{3\pi \mel c^3 \rblr^2}.
\eqe
Thus, the total photon compactness in the emission region, which is relevant to the calculations of photohadronic emission,
is given by $\ell_{\rm tot} = \lblr + \ell_{\rm syn}$, 
where $\ell_{\rm syn}$ is the compactness of internally produced synchrotron photons. We note that we do not take into account 
the anisotropy of the BLR photon field as seen in the emission region in the calculations of photopion production (see also, \citealt{atoyandermer01, 
tavecchioetal14}).
\subsection{Method}
\label{method}
The algorithm we follow in our numerical approach is described below.
\subsubsection{No external radiation}
\label{method-1}
We start by assuming that $\lblr$ is negligible with respect to the 
compactness of the internally produced photons, namely $\lblr \ll \ell_{\rm syn}$.
This can be seen as the case of minimal compactness. For a given pair of $B$ and $R$ we 
\begin{enumerate}[]
 \item choose a high value for the Doppler factor $\delta$, e.g. 50;
 \item choose values for the rest of the parameters, e.g. $\ell_{\rm p,e}$ and $\gpmx$,
  that lead to a reasonable fit of the SED. {This is defined by the curve}
   that passes within the error 
  bars of most of the observational points, with particular 
  emphasis on the highest energy ones, that are more 
  directly affected by secondary particles from photohadronic 
  processes within this model. We are not interested in the 
  absolute best fit, as would be determined by a $\chi^2$ test, but rather a good enough one, as determined visually;  
  \item if the derived photon spectrum describes the SED reasonably well, we return to
 step (i) and choose a smaller value of $\delta$; if the photon spectrum does not
 fit the SED for the adopted Doppler factor because of enhanced photohadronic emission, we stop and
 define the current value of the Doppler factor  as $\delta_{\min}$.
 \end{enumerate}
 \subsubsection{Internal and external radiation}
 \label{method-2}
As a second step, we include the BLR emission into the calculation of the broad-band photon spectrum.
For each value of $\delta \approx \Gamma$,  
we obtain the multi-wavelength spectra for different values of $\lblr$. 
Each pair of $(\delta, \lblr)$ translates into a pair of $(\delta, x)$ through eq.~(\ref{lblr}).
Thus, numerical runs for fixed $\delta$ and $\lblr$ imply 
different {locations} of the emission region in the jet, i.e. 
\eqb
x = \left(\frac{\ell_0}{\lblr} \delta^2-1\right)^{1/4}.
\label{x}
\eqe
This corresponds also to different comoving photon energies $\eblr'$ (see eq.~(\ref{eblr})).
It is important to note that the inclusion of the BLR radiation does not
affect the values of $\ell_{\rm e, p}$ and $\gpmx$ that we derived previously (Sect.~\ref{method-1}), 
since the SED is fitted by the synchrotron radiation
of primary electrons and protons. 
This shows the secondary role of the external radiation in the proton synchrotron model,
in contrast to the EC leptonic models.

We then determine that value of $\lblr$ above which the 
cascade emission modifies the proton synchrotron radiation spectrum {at a few GeV} in a way that 
{the total emission exceeds the} observations. As can be evidenced by eq.~(\ref{x}), the
maximum value of the BLR compactness translates into a lower limit of  $r/\rblr$.
This parametrization of the problem allows, therefore, for solutions within or outside the BLR and is a generalization of the approach
presented in Sect.~\ref{method-1}.
We note that the cascade emission depends on both $\lblr$ and $\eblr'$, which affect the photohadronic production rates
in a direct and indirect way, respectively. The value of $\eblr'$ affects the energy thresholds
for photohadronic interactions.  Thus, for a given Doppler factor, a choice
of a higher value of $\lblr$ does not necessarily mean that the luminosity of the cascade emission will be higher.

\section{Results: application to 3C~273}
\label{results}
We apply our method to the well-known FSRQ 3C~273 at redshift $z=0.158$.
The optical-UV spectrum of 3C~273 shows a
prominent excess of emission, which is mainly interpreted as
a contribution of the accretion disk emission (see Ulrich, 1981;
Soldi et al., 2008, and references therein). The detection of lines
in the optical-UV spectrum of 3C 273, e.g. Ly-$\alpha$, CIV, OVI, CIII, NIII, and SVI  \citep[e.g.][and references
therein]{paltani03} is connected with the BLR.
The accretion disk and BLR luminosities of 3C~273 are well-defined, i.e.
$L_{\rm ad}=1.3\times 10^{47}$~erg/s \citep{vasudevan09} and $L_{\rm BLR}=9.1\times 10^{45}$~erg/s \citep{peterson04}. 
The knowledge of both $L_{\rm ad}$ and $L_{\rm BLR}$ {reduces} 
the number of free parameters entering in the model \citep[see also][]{boettcherreimer13}. 
Other parameters describing the BLR are $\rblr \approx 1.1 \times 10^{18}$~cm (from eq.~(\ref{rblr})), $\eblr=8$~eV and $\xi_{\rm BLR}=0.1$.
As can be evidenced by the ASI Science Data Centre (ASDC)\footnote{http://www.asdc.asi.it/SED}, there 
is a huge amount of archival and non-simultaneous observations for 3C~273. {As simultaneous multi-wavelength observations are important for our analysis, 
we use the dataset by \cite{abdo10_sed}. We emphasize, though, that our method
can be easily applied to different broad-band simultaneous data, since
it is based on a generic idea.

\begin{figure}
 \centering
 \includegraphics[width=0.5\textwidth]{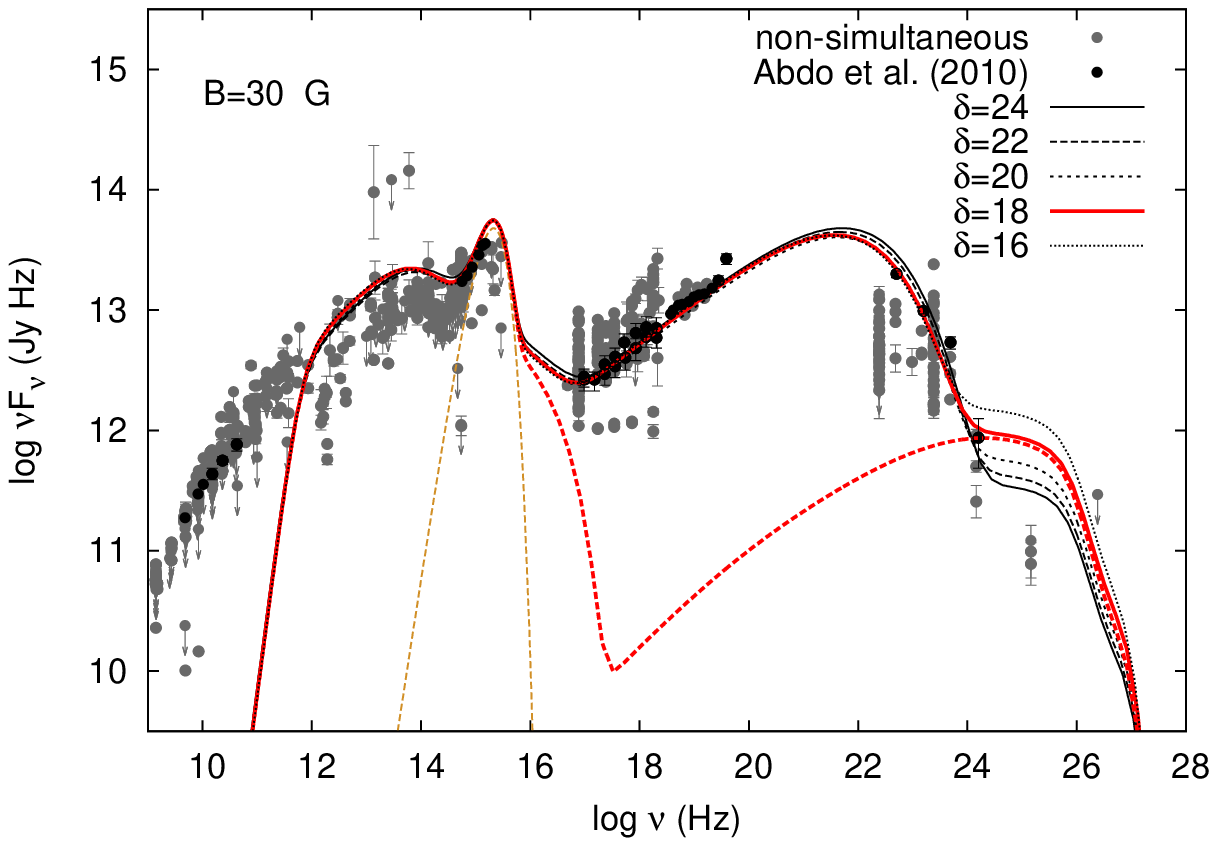}
 \includegraphics[width=0.5\textwidth]{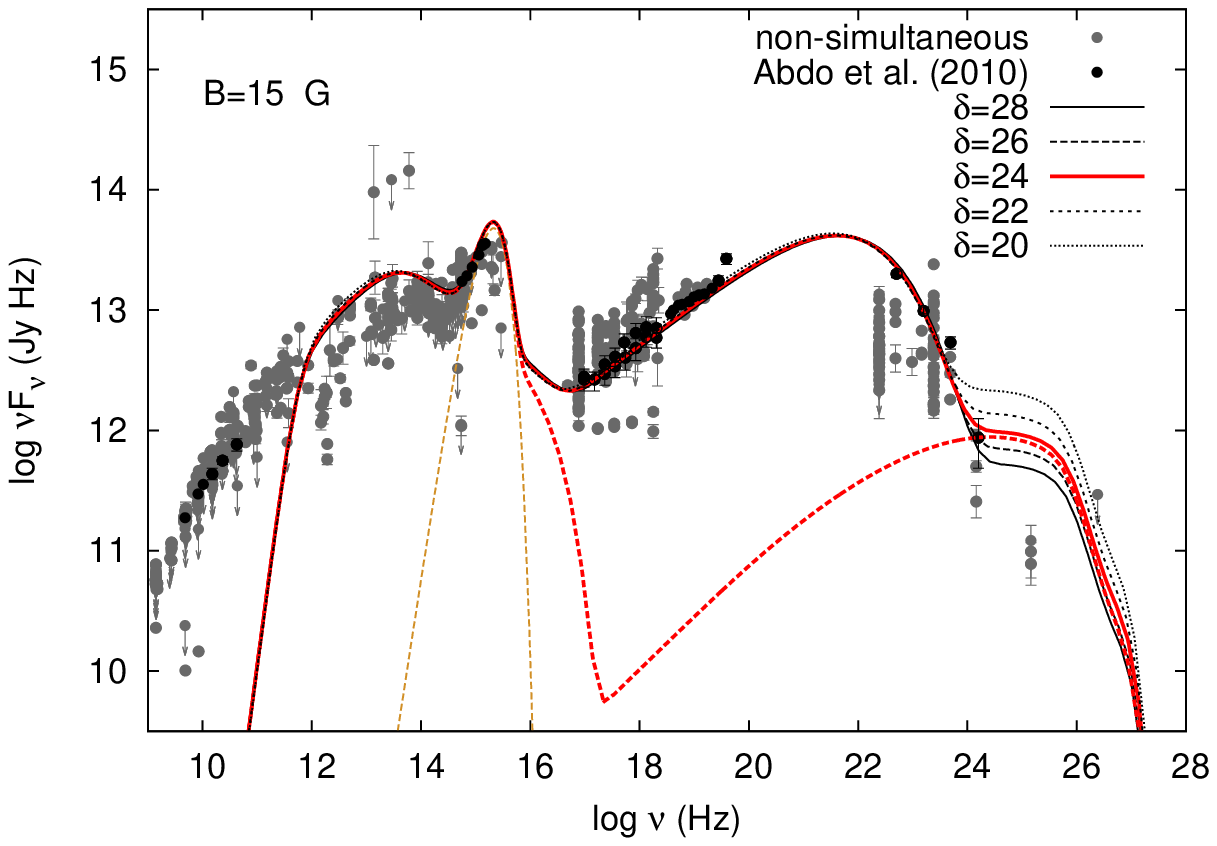}
 \includegraphics[width=0.5\textwidth]{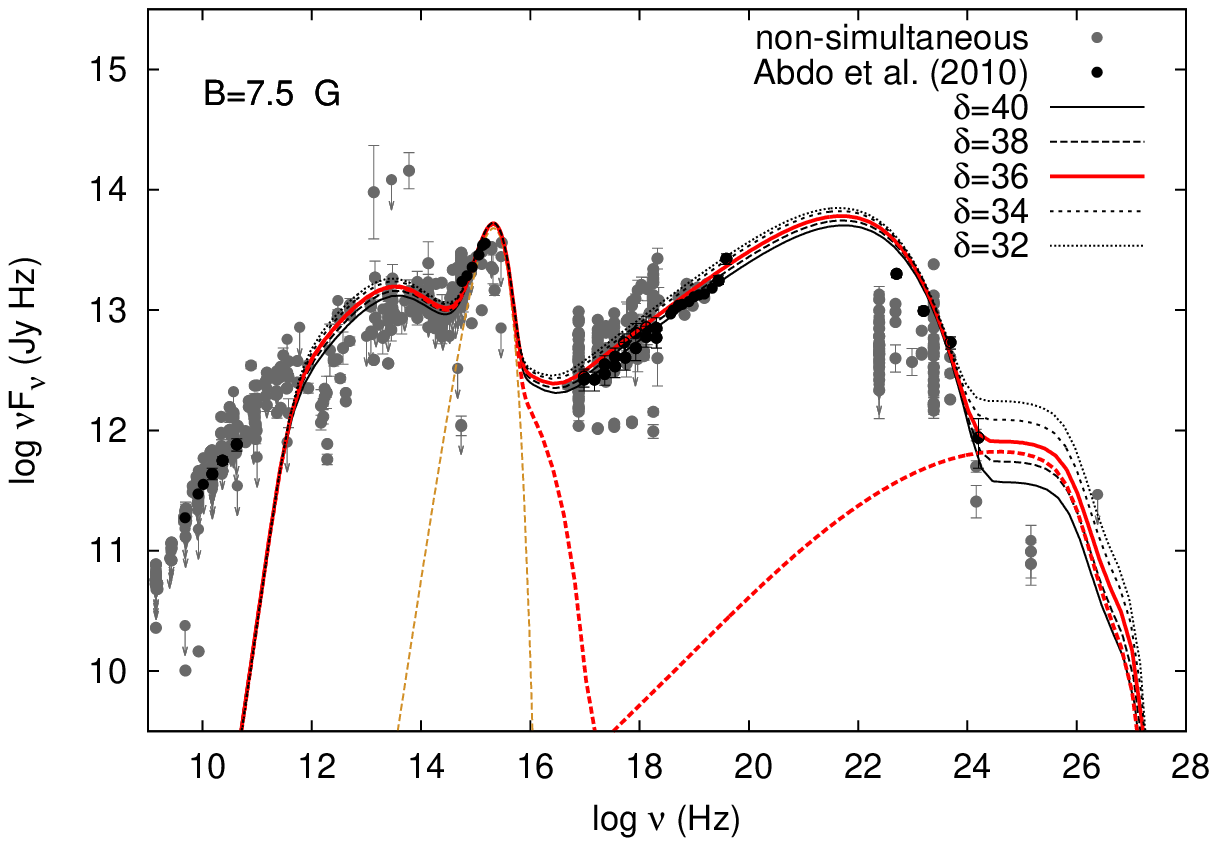}
 \caption{Proton synchrotron model fits to the simultaneous multi-wavelength data of 3C~273 
 by \citealt{abdo10_sed} (black symbols) for B=30 G, 15 G and 7.5 G (top to bottom),
 in the limiting case of $\lblr \ll \lsyn$ {or, equivalently $r \gg \rblr$}.
 In each panel, the SED is modelled using different values of the Doppler factor. Spectra
 shown with red lines correspond to the minimum value of the Doppler factor that can
explain the simultaneous SED. The respective spectrum when proton synchrotron radiation is omitted, which reveals the underlying spectrum
of the EM cascade, is shown with a red dashed line. The accretion disk spectrum is overplotted with an orange dashed line.
Grey symbols are non-simultaneous and archival observations taken from ASDC.}
 \label{fit_noext}
\end{figure}
\subsection{No external radiation or $r \gg \rblr$}
Figure \ref{fit_noext} demonstrates the effect of the cascade emission on the high-energy part of the spectrum as the Doppler factor
of the emission region progressively decreases, in the minimal scenario where 
only internal radiation is a target for photohadronic interactions.  The three panels (from top to bottom)
correspond to different magnetic field strengths, namely B=30 G, 15 G and 7.5 G. {The radius
of the emission region is assumed to be $R=3.6\times 10^{16}$~cm in all runs. For a fiducial value of $\delta=15$, 
our choice results in $t_{\rm var}\sim 1$~day, which is typical for 3C~273 \citep{courvoisier88}.}
Other parameters used and kept fixed in the numerical runs are:
$\gamma_{\rm e,\min}=1.6\times 10^2$, $\gamma_{\rm p, \min}=1$, $\gamma_{\rm e, \max}=5\times 10^3$,
$p_{\rm e}=2.7$ and $p=2.3$; both distributions of primary particles were modelled as 
$n_i \propto \gamma^{-p_{\rm i}} e^{-\gamma/\gamma_{\rm i, \max}}$, $i=e,p$.
The parameters that had to be adjusted in order to model the SED for the different
Doppler factor values are listed in Table~\ref{tab-0}.
\begin{table}
\centering
\caption{Parameter values used for modelling the multi-wavelength emission of 3C~273, as illustrated in Fig.~\ref{fit_noext}. 
Not all cases listed in this Table are depicted in Fig.~\ref{fit_noext}, but we include them for completeness reasons.
Other parameters used are kept fixed (see text).}
\begin{tabular}{cccc}
\hline
    & $\delta$ & $\ell_{\rm e}$ (in log) & $\ell_{\rm p}$ (in log) \\
    \hline \hline
B=30 G & \multicolumn{3}{c}{\phantom{}} \\
$\gamma_{\rm p, \max}=10^8$ & \multicolumn{3}{c}{\phantom{}} \\
\hline
			     &  28 & -5.2 & -3.5 \\
			     &  26 & -5.1 & -3.4 \\
			     &  24 & -4.9 & -3.2 \\
			     & 22 & -4.8 &  -3.1 \\
			     & 20 & -4.6 &  -3.0 \\
($\dmin$)		     & 18 & -4.4 &  -2.8 \\
			     & 16 & -4.2 &  -2.6 \\
\hline			     
B=15 G & \multicolumn{3}{c}{\phantom{}} \\
$\gamma_{\rm p, \max}=1.2\times 10^8$ & \multicolumn{3}{c}{\phantom{}} \\
\hline
			      &  32 & -5.5& 	-3.3\\
			      &  30 & -5.3& 	-3.1\\
			      &  28 & -5.2& 	-3.0\\
			     & 26 & -5.1 &  -2.9 \\
($\dmin$)		     & 24 & -4.9 & -2.7\\
			     & 22 & -4.8 & -2.6 \\
			     & 20 & -4.6 & -2.4 \\
\hline
B=7.5 G & \multicolumn{3}{c}{\phantom{}} \\
$\gamma_{\rm p, \max}=1.6\times 10^8$ & \multicolumn{3}{c}{\phantom{}} \\
\hline
			      &  40 & -6.0 &-3.0\\
			     & 38 & -5.9& -2.9\\
($\dmin$)		     & 36 & -5.7& -2.7\\
			     & 34 & -5.6& -2.6\\
			     & 32 & -5.5& -2.5 \\
\hline
 \end{tabular}
\label{tab-0}
 \end{table}
In all cases, the plateau-like emission above a few GeV ($> 10^{24}$~Hz) is the result
of the EM cascade initiated by VHE $\gamma$-rays produced in photohadronic interactions. 
Spectra shown with red thick lines correspond to the minimum value of the Doppler factor that can
explain the simultaneous SED. For $\delta < \dmin$, the photon spectra above $>10^{24}$~Hz exceed the observations.

At this point, it is interesting to compare 
the  numerically derived $\ell_{\rm p}$ and $\dmin$ listed in Table~\ref{tab-0} with the respective values
predicted by our analysis in Sect.~\ref{analytical}. This is exemplified in Fig.~\ref{fig-comparison} where the top and bottom panels show
the comparison for $\ell_{\rm p}$ and $\dmin$, respectively. {In both panels,} the values from the numerical
analysis are shown with symbols, while the curves are calculated using eq.~(\ref{lp}) for $R=3.6\times 10^{16}$~cm,
$\nu_{\gamma}=10^{22}$~Hz, $L_{\gamma} = 6.3\times 10^{46}$~erg/s and $p=2.3$ {(top panel)} and 
eq.~(\ref{dmin1}) for $\nu_{\rm s} = \es/h = 3.2\times 10^{13}$~Hz, 
$L_{\rm s}=6.3\times10^{45}$~erg/s , $\nu_{\gamma}=10^{22}$~Hz, $\eta=0.2$, $p=2.3$, 
$\blow=0.7$, and $\bhigh=1.35$ {(bottom panel)}.

In both panels, our analytical curves are in good agreement
with the numerical values determined through the SED modelling, with some deviation
becoming systematically larger
for $B=7.5$~G (top panel) and $R=3.6\times10^{15}$~cm (bottom panel). 
However, it is remarkable how well the analytical curves follow the trend found numerically, and 
especially for $\dmin$, since we made several approximations in order to derive an analytical expression (eq.~\ref{dmin1}).
A quantitative difference between the curves is something to be expected, since the
 numerical analysis: (i) takes into account
the emission from secondary pairs (from Bethe - Heitler and $\pi^{+}$ decays) in the formation of the EM cascade,
(ii) makes no assumptions about the photohadronic production rates, and (iii) takes into account
the spectral shape of the EM cascade.

\begin{figure}
 \centering
 \includegraphics[width=0.5\textwidth]{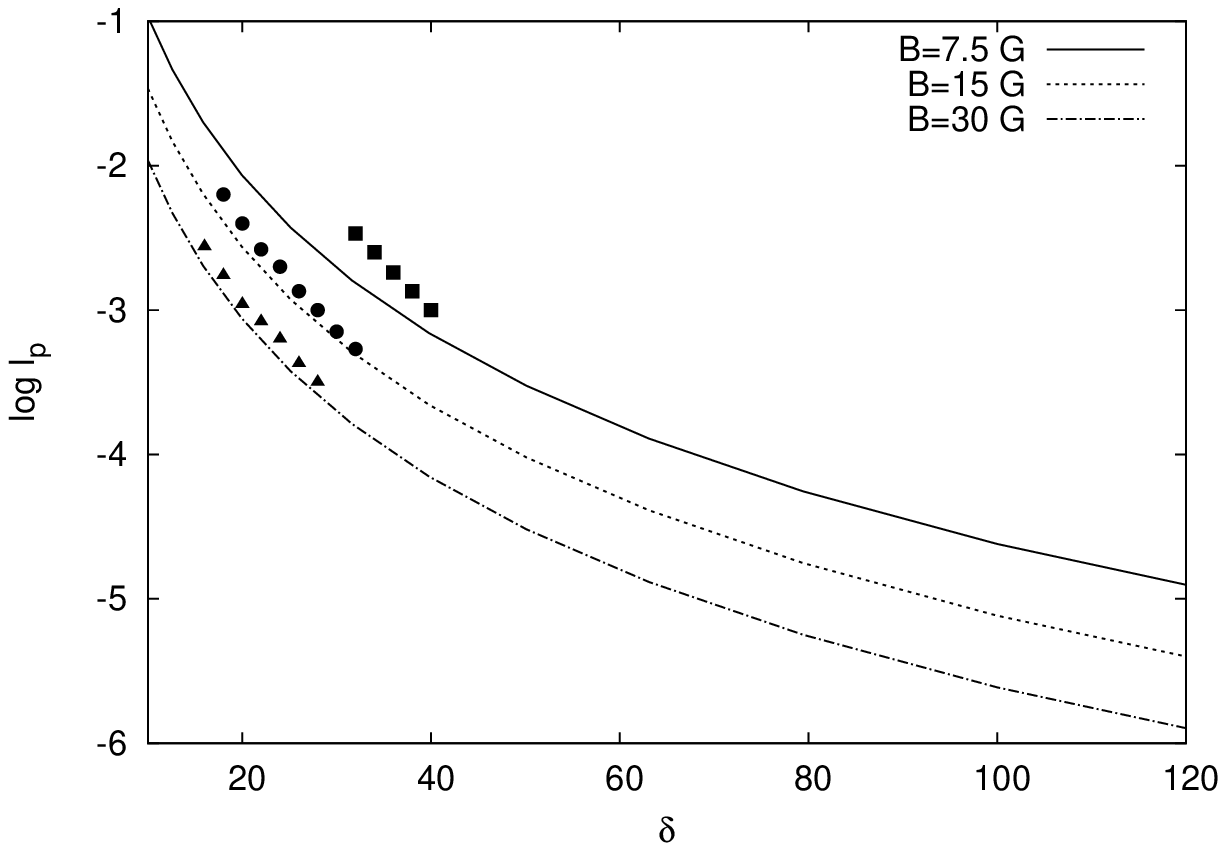}
  \includegraphics[width=0.5\textwidth]{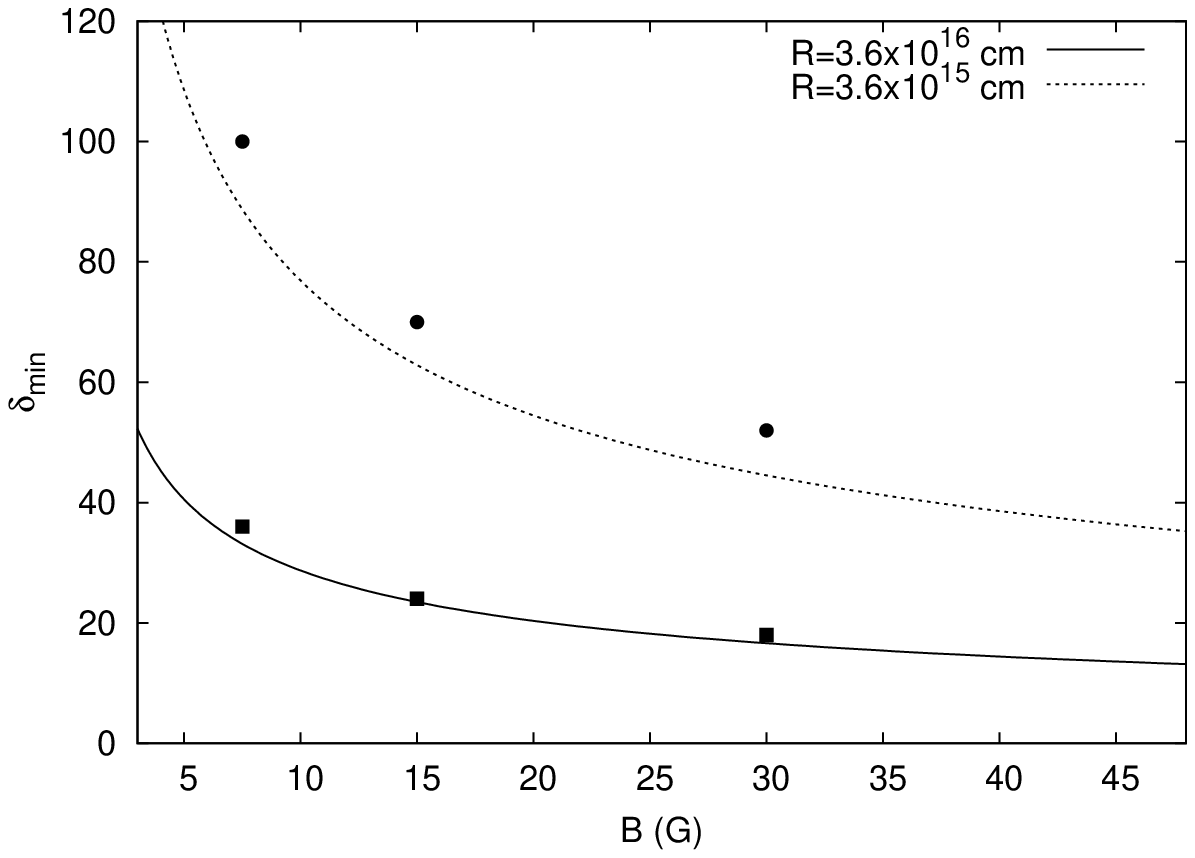}
\caption{Comparison between the analytical ({curves}) and numerical
(symbols) results. Top panel: $\ell_{\rm p}$ as a function of $\delta$ 
{for 
$B=7.5$~G (solid line/squares), 15~G (dotted line/circles) and 30~G (dashed-dotted line/triangles).}
Bottom panel: $\dmin$ as a function of $B$ (bottom panel) for {$R=3.6\times 10^{16}$~cm (solid line/squares)
and $R=3.6\times 10^{15}$~cm (dotted line/circles).}
}
\label{fig-comparison}
\end{figure}

For completeness reasons, we repeated the modelling procedure for a smaller emission region having $R=3.6\times 10^{15}$~cm.
The bottom panel of Fig.~\ref{fig-comparison} shows that unreasonably high values of the Doppler factor ($\delta \gtrsim 40$) are required, 
in this case, to avoid
the effects of the EM cascade, while only very strong magnetic fields ($B \gg 50$~G), can bring the Doppler factor
to lower values.

\subsection{Internal and external radiation}
Following the method described in Sect.~\ref{method-2}, we derived the minimum
value of the ratio $r/\rblr$ for the three cases considered previously.
Our results are summarized in Table~\ref{tab-1} and Fig.~\ref{xmin}. 
\begin{figure}
 \centering
 \includegraphics[width=0.5\textwidth]{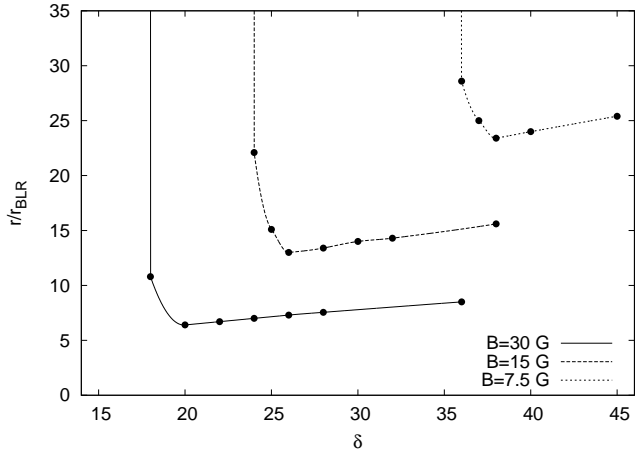}
\caption{Plot of the ratio $r/\rblr$ as a function of the Doppler factor for three values
of the magnetic field marked on the plot and $R=3.6\times 10^{16}$~cm. Symbols and lines are used
for displaying the numerical results and the interpolated values, respectively. Only the region above each curve
is allowed, as values below the curve lead to significant emission from the EM cascade.
The abrupt increase of $r/\rblr$ occurs at $\delta=\dmin$ {defined by eq.~(\ref{dmin1})}.} 
\label{xmin}
\end{figure}
\begin{table}
\centering
\caption{Maximum  $\lblr$ and, equivalently, minimum $r/\rblr$ as determined by modelling 
of the multi-wavelength emission of 3C~273 for a given Doppler factor and magnetic field strength.
The method for the determination or their values is described in Sect.~\ref{method-2}.
All other parameters are same as in Table~\ref{tab-0}.}
\begin{tabular}{cccc}
\hline
    & $\delta$ & $\lblr$ & $r/\rblr$  \\
    \hline \hline
B=30 G & \multicolumn{3}{c}{\phantom{}} \\
\hline
			     & 28 & $10^{-4}$ & 7.5  \\
			     & 26 & $10^{-4}$ & 7.3 \\
			     & 24 & $10^{-4}$ & 7.0 \\
			      & 22 & $10^{-4}$ & 6.7 \\
			     & 20 & $10^{-4}$ & 6.4 \\
			     & 18 & $10^{-5}$ & 10.8 \\
\hline			     
B=15 G & \multicolumn{3}{c}{\phantom{}} \\
\hline
			      & 38 & $10^{-5}$& 15.6\\
			     & 32 & $10^{-5}$ & 14.3 \\
			      & 30 & $10^{-5}$ & 13.9\\
			     & 28 & $10^{-5}$& 13.4 \\
			     & 26 & $10^{-5}$& 12.9 \\
			     & 25 & $5\times 10^{-6}$& 15.1 \\
			     & 24 & $10^{-6}$& 22.1 \\
\hline
B=7.5 G & \multicolumn{3}{c}{\phantom{}} \\
\hline
			    &  45 & $2\times 10^{-6}$ &25.4\\
			     & 40 & $2\times 10^{-6}$& 24.0\\
			      & 38 &  $2\times 10^{-6}$& 23.4\\
			     & 37 &  $10^{-6}$& 27.4\\
			     & 36 &  $8\times 10^{-7}$& 28.6\\
\hline
 \end{tabular}
\label{tab-1}
 \end{table}
The different
curves are obtained for $B=30$~G, 15~G and 7.5~G. The results obtained from the numerical
fitting are shown as symbols, while the curves are the result of interpolation.  Only the region above each curve
is allowed, as values below the curve lead to significant emission from the {EM cascade}. Thus, each curve
is the locus of points corresponding to the minimum distance, $r_{\min}$, for various Doppler factor values.
The curves can be also safely extrapolated to higher Doppler factor values, 
since a power-law dependence is established, namely $r_{\min}/\rblr \propto \delta ^{1/2}$.
The abrupt increase of $r_{\min}/\rblr$ occurs at $\delta=\dmin$, as expected. We remind that $\dmin$ 
is derived in the limiting case of $\lblr \ll \lsyn$ or, equivalently, $r \gg \rblr$.
 
Figure \ref{xmin} reveals the following trend: curves move from the lower left part to the upper right
part of the plot for progressively weaker magnetic fields. 
The horizontal shifting of the curves can be easily understood by inspection of eq.~(\ref{dmin1}), which shows  that
$\dmin \propto B^{-(3+\bhigh)/({p+3}+3\bhigh)}$. For a given $\delta$, weaker magnetic fields require higher values of the proton compactness
to explain the observed $\gamma$-ray luminosity, which in turn enhances the EM cascade
emission. To avoid an {excess in $\gamma$-rays due to the cascade emission, }
the emission region should be located  even further out. 
This qualitatively explains the vertical shift of the curves in Fig.~\ref{xmin}.

Figure~\ref{xmin} shows that
the emission region of 3C~273 in the proton synchrotron scenario cannot be located in the BLR region, at least
for $B\le 30$~G and $R=3.6\times 10^{16}$~cm. Let us discuss how a different
choice of $R$ and $B$ would affect our conclusion.
A choice of a smaller radius, e.g. $R\sim 10^{15}$~cm, would
have a similar effect on the curves as that of a decreasing magnetic field (shifting to the upper right part of Fig.~\ref{xmin}).
Only if the emission region were larger and, thus, less compact could it be located within the BLR.
However, as we show in Sect.~\ref{constraints}, this scenario becomes less plausible when
the variability of the source is taken into account. In addition, the jet power of a larger emission region
would significantly exceed the accretion power
of 3C~273 (see details in Sect.~\ref{constraints}). 

In principle,
the emission region could
be located within the BLR for sufficiently large magnetic fields, namely $B \gg 30$~G, according to the trend
we find in Fig.~\ref{xmin}. The question that arises in this case is whether the required $B$ values are plausible or not.
Instead of performing additional simulations, we can address the question with the analytical tools presented
in Sect.~\ref{analytical}. Inspection of Tables~\ref{tab-0} and \ref{tab-1}, shows
that at the minimum distance $r_{\min}$ the electron and BLR compactnesses are approximately equal, while
$\delta = \dmin {+ \epsilon}$,
with $\epsilon\simeq 1-2$. Since the electron compactness is a good proxy
for the internal synchrotron photon compactness, 
we can estimate the minimum distance by requiring $\lblr(\dmin) \simeq \lsyn(\dmin)$ or, equivalently
\eqb
\frac{\dmin^2 \ell_0}{1+x_{\rm min}^4} \simeq \frac{\sth L_{\rm syn}}{4\pi R\mel c^3 \dmin^4},
\eqe
where $\dmin$ is given by eq.~(\ref{dmin1}). For $R=3.6\times 10^{16}$~cm and
for all other parameters same as in Fig.~\ref{fig1}, we find that $r_{\min}/\rblr < 1$ for $B\gtrsim 450$~G.
We can therefore argue that the emission region of 3C~273 cannot be located within the BLR 
for plausible parameter values.


\subsection{Additional constraints}
\label{constraints}
The constraints on the Doppler factor and on the
distance of the emission region from the super-massive black hole were derived 
based only on the photohadronic emission and the initiated EM cascade.
These can become even more tight when combined with information about the variability of the source and the energetics of the emission region. 

\subsubsection{Variability}
The observed high-energy variability in blazars may range from hours up to few days depending on the flaring activity.
For example, the shortest variability timescales probed by FERMI-LAT are several hours \citep{tavecchioetal10, foschinietal11, saito13}.
In order to keep our analysis as generic as possible,
we chose $R$ and $\delta$ to be independent parameters. Here, we revise our previous results 
by including the variability information. If $\tvar$ is the observed variability timescale, $\delta$ and $R$ 
should satisfy the  causality condition
\eqb
\delta \ge \delta_{\rm var}\equiv \frac{R(1+z)}{c\tvar}.
\label{dvar}
\eqe
Figure \ref{fig2} illustrates the revised parameter space $\delta-B$ for $R=3.6\times 10^{16}$~cm and three indicative values
of the observed variability timescale marked on the plot. The regions that lie above
the horizontal dotted and solid lines denote areas {where} both the EM cascade and variability
constraints are satisfied. {For $\tvar=1$~d, the variability offers no additional constraint over $\dmin$.
However, for $\tvar \lesssim 12$~h the lower limit
$\delta_{\rm var}$ becomes more constraining than the lower limit derived imposed by the EM
cascade.}

\begin{figure}
 \centering
 \includegraphics[width=0.5\textwidth]{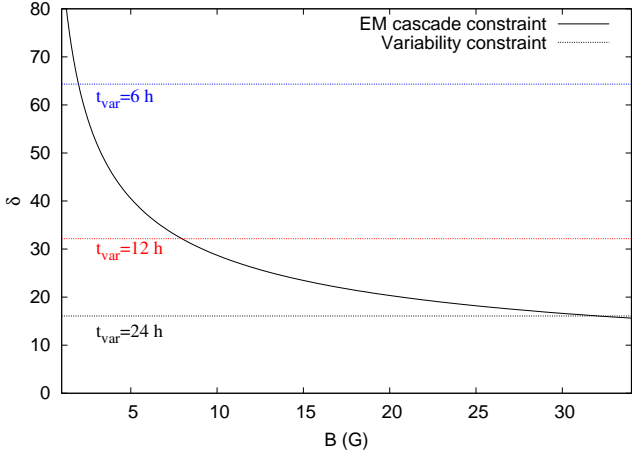}
\caption{Same as Fig.~\ref{fig1} with the addition of the variability constraint $\delta \ge \delta_{\rm var}$ (dotted lines) 
for $R=3.6\times 10^{16}$~cm and three indicative values of $\tvar$ marked on the plot. 
The curves of $\dmin$ are calculated using eq.~(\ref{dmin1}). Only the regions that lie above
the horizontal dotted lines, satisfy the variability constraint. All other parameters are same
as in Fig.~\ref{fig1}.} 
\label{fig2}
\end{figure}
\begin{figure}
 \centering
 \includegraphics[width=0.5\textwidth]{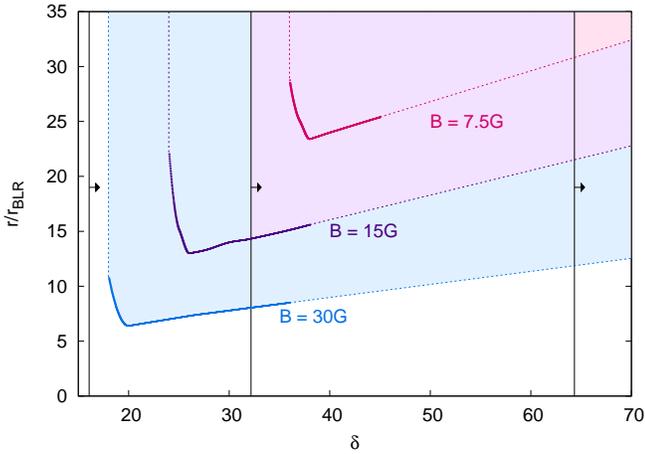}
\caption{The revised parameter space $r/\rblr - \delta$ for 3C~273 after taking into account
the variability constraint $\delta \ge \delta_{\rm var}$ for  $R=3.6\times 10^{16}$~cm and three values of the magnetic field 
marked on the plot. Vertical lines denote $\delta_{\rm var}$ given by eq.~(\ref{dvar}). From left to right: $t_{\rm var}=24$~h, 12~h and 6~h.
The coloured regions above the curves and on the right of the vertical lines denote areas where both the variability 
and EM cascade constraints are satisfied. All other parameters are same
as in Fig.~\ref{fig1}.} 
\label{xmin2}
\end{figure}

{Similarly, }the $r/\rblr$-$\delta$ parameter space can be further constrained by including the variability constraint.
{This is illustrated in Fig.~\ref{xmin2}.}
The vertical lines show $\delta_{\rm var}$ calculated using eq.~(\ref{dvar}) for $R=3.6\times 10^{16}$~cm and three
indicative values of $t_{\rm var}$ (from left to right,  $t_{\rm var}=24$~h, 12~h and 6~h). 
To the right of those vertical lines are regions where the causality condition
is satisfied. The EM cascade does not exceed the
$\gamma$-ray  observations {at a few GeV} for parameters drawn above the curves. Finally, the coloured regions denote the parameter space
where both requirements are satisfied. We find that for $\sim$day timescale variability the location of the emitting
region and the Doppler factor are only limited by the EM cascade. A shorter variability timescale, which has been observed
during bright $\gamma$-ray flares of 3C~273 \citep[e.g.][]{rani13}, can impose tighter constraints on the minimum $r$ and $\delta$; 
{e.g., for $\tvar=6$~h and $B=7.5$~G only the upper right corner of the parameter space is allowed.}

\subsubsection{Jet power}
In FSRQs {the accretion disk luminosity can be estimated
using the BLR luminosity, and the accretion power ($\pacc=\dot{M}c^2$) can be then calculated as
\citep[e.g.][]{ghisellini14}
\eqb
\pacc = 10 \frac{L_{\rm ad} }{\epsilon_{\rm r,-1}},
\label{pacc}
\eqe
{where $\epsilon_{\rm r}$ is the radiative efficiency.}
In general, the jet power is written as $\pjet=\epsilon_{\rm j} \pacc$, with
$\epsilon_{\rm j}\lesssim 1.5$ \citep[][and references therein]{zdziarskiboettcher15}. Although 
the jet power can exceed the accretion one \citep[e.g.][]{tchekhovskoy11}  due
to the efficient extraction of energy from a Kerr black hole \citep[][]{BZ77}, here we consider
the more conservative case of $\epsilon_{\rm j}=1$. 
We therefore} impose
the following `energetic' constraint 
\eqb
\pjet \le \pacc,
\label{energetic}
\eqe
 Neglecting the cold proton  and radiation energy densities, the power of a two-sided jet 
 can be written as \citep[e.g.][]{ghisellini14}
\eqb
\pjet \approx 2\pi R^2 \Gamma^2 c \left(u'_{\rm e}+u'_{\rm p} + u'_{\rm B}\right)
\eqe 
where $u'_i$ ($i=e,p,B$) is the energy density as measured in the rest frame of the emission region.
Dropping the electron term (see also Table~\ref{tab-2}) and using eq.~(\ref{lp}) with $\delta \approx \Gamma$,
we write the jet power as
\eqb
\pjet \approx \frac{R^2 c}{4}\left[A \left(B \delta \right)^{-(p+1)/2} + \left(B\delta \right)^2 \right],
\label{pjet}
\eqe
where 
\eqb
A= \frac{6C_{\rm p}\mpr c^2 L_{\gamma}}{f_{\rm p} R^3 \nu_{\gamma}^{(3-p)/2}}.
\eqe
The jet power given by eq.~(\ref{pjet}) for $R=3.6\times 10^{16}$~cm, $L_{\gamma}=6.3\times 10^{46}$~erg/s,
$\nu_{\gamma}=10^{22}$~Hz, $p=2$, and three values of the magnetic field, i.e. $B=7.5$, 15 and 30~G,
is plotted as a function of $\delta$ in Fig.~\ref{power}. Overplotted with symbols
are the values calculated using eq.~(\ref{pjet}) for $u'_{\rm p}$ determined by 
the numerical SED modelling of 3C~273 (see also Table~\ref{tab-2}).
Apart from an offset ($\lesssim 3$) between the analytical curve and the numerical
values for the case of $B=7.5$~G (see also Fig.~\ref{fig-comparison}), the two results
are in good agreement. 
\begin{figure}
 \centering
 \includegraphics[width=0.49\textwidth]{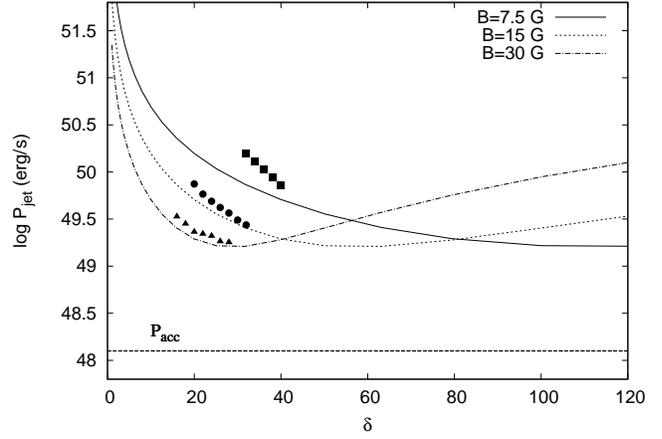}
\caption{Jet power, as derived in the proton synchrotron model for 3C~273. The values
derived by numerically modelling the SED of 3C~273 (see Table~\ref{tab-2}) are shown as symbols, while
the {curves} are calculated using eq.~(\ref{pjet}) {for 
$B=7.5$~G (solid line/squares), 15~G (dotted line/circles) and 30~G (dashed-dotted line/triangles).}
The horizontal dashed line marks the accretion luminosity given by eq.~(\ref{pacc}). } 
\label{power}
\end{figure}
Figure~\ref{power} shows that in all our SED fits the jet power is dominated by the energy density of
relativistic protons, i.e. $P_{\rm jet} \propto \left(B \delta \right)^{-(p+1)/2}$,
while it exceeds  $\pacc =1.3 \times 10^{48}$~erg/s (dashed line).
Even the minimum jet power ($P_{\rm jet,\min}$) exceeds the accretion power 
by approximately one order of magnitude \citep[see also][]{zdziarskiboettcher15}. 
Thus, at least for the particular choice of $R$ and $B$, 
{relation (\ref{energetic}) cannot be satisfied, and in this regard, it cannot further
constrain the parameter space.}

\begin{table}
\centering
\caption{Energy densities of relativistic electrons and protons (in the comoving frame)
as determined by modelling 
of the multi-wavelength emission of 3C~273 for a given Doppler factor and magnetic field strength.
The jet power is listed in the last column of the table.
All other parameters are same as in Table~\ref{tab-0}.}
\begin{threeparttable}
\begin{tabular}{ccccc}
\hline
    & $\delta$ & $u'_{\rm e}$\tnote{a} & $u'_{\rm p}$\tnote{b} & $P_{\rm jet}$\tnote{c} \\
    \hline \hline
B=30 G & \multicolumn{4}{c}{\phantom{}} \\ 
($u'_{\rm B}=35.8$~erg/cm$^3$) & \multicolumn{3}{c}{\phantom{}}\\
\hline
			     & 28 & 1.7 & 58 & 1.8\\
			     & 26 & 2.3  & 75 & 1.8\\
			     & 24 & 3.3 & 113& 2.1 \\
			      & 22 & 4.4  & 150 & 2.2 \\
			     & 20 & 6.5 & 200 & 2.3\\
			     & 18 & 11.0 & 320 & 2.8\\
			     & 16 & 16.8 & 502 & 3.3\\
\hline			     
B=15 G & \multicolumn{4}{c}{\phantom{}} \\
($u'_{\rm B}=9$~erg/cm$^3$) & \multicolumn{3}{c}{\phantom{}}\\
\hline

			     & 32 & 2.8& 100 & 2.7 \\
			     & 30 & 3.8& 132 & 3.1 \\
			     & 28 &5.1& 182 & 3.7\\
			     & 26 & 6.8 & 245 & 4.2\\
			      & 24 &9.6& 339 & 4.9\\
			        & 22 &13.7& 483 & 5.8\\
			          & 20 &20.5& 753 & 7.4\\
\hline
B=7.5 G & \multicolumn{4}{c}{\phantom{}} \\
($u'_{\rm B}=2.2$~erg/cm$^3$)  & \multicolumn{4}{c}{\phantom{}} \\
\hline
			    &  40 & 2.2 &182& 7.2\\
			     & 38 & 3.0& 244& 8.7 \\
			      & 36 &  4.1& 332& 10.6 \\
			     & 34 &  5.5 & 458&13.0 \\
			     & 32 & 7.5 & 621& 15.6\\
\hline
 \end{tabular}
 \tnote{a} Electron energy density in units of $10^{-5}$ erg/cm$^3$.\\
 \tnote{c} Electron energy density in erg/cm$^3$. \\
 \tnote{b} Jet power in units of $10^{49}$~erg/s.
  \end{threeparttable}
\label{tab-2}
 \end{table}
 
The relatively good agreement between the analytical and numerical results for the jet power allows
us to use expression (\ref{pjet}) for investigating the dependence of $P_{\rm jet,\min}$ on
the parameters, and to search for those, if any, that can bring the jet power
closer to the accretion luminosity. 
The jet power given by eq.~(\ref{pjet}) is minimized for
\eqb
\delta_0 B_0 = \left(A\frac{p+1}{4}\right)^{2/(p+5)}
\label{min-condition}
\eqe
and its minimum value for a given source ($L_{\gamma}, \nu_{\gamma}$ and $p$, fixed) depends only the radius
$R$ through
\eqb
P_{\rm jet, \min} = \tilde{A} R^{(2p-2)/(p+5)},
\label{pmin}
\eqe
where 
\eqb
\tilde{A}= \frac{c}{4}\left(\frac{6C_{\rm p}\mpr c^2 L_{\gamma}}{f_{\rm p}\nu_{\gamma}^{(3-p)/2}}\right)^{4/(p+5)} \left(s^{-(p+1)/(p+5)}+s^{4/(p+5)}\right)
\eqe
and 
\eqb
 s = \frac{p+1}{4}.
\eqe
The requirement $P_{\rm jet, \min}= \pacc$ is satisfied for
\eqb
R_0  =\left( \frac{\pacc}{\tilde{A}}\right)^{(p+5)/(2p-2)} 
\label{Ro}
\eqe
which for 3C~273 and $p=2.3$ becomes $R_0\sim 10^{14}$~cm. Substitution of 
$R_0$ into eq.~(\ref{min-condition}) results in $\delta_0 B_0 \sim 5\times 10^4$ (in cgs units), namely
the Doppler factor and the magnetic field should take extreme values.

Summarizing, we showed explicitly that in the proton synchrotron scenario for the multi-wavelength emission
of 3C~273 there are no reasonable physical parameters that can bring the jet power close the accretion power.
In all cases, we find $\pjet \gtrsim 10\pacc$, in agreement with the independent analysis by \cite{zdziarskiboettcher15}.

\section{Discussion}
\label{discuss}
The derivation of constraints for the Doppler factor and/or the location of 
the emission region in the leptonic framework of blazar 
emission has been the subject of 
several studies 
\citep[e.g.][]{Dondi1995, poutanenstern10, tavecchioetal10, dotson12, cerrutietal13, dermeretal14, nalewajko14, zacharias15}.
The exploration of such constraints within the confines of alternative models is a method
that may lead to their eventual verification or exclusion, 
with new observations potentially reshaping the available parameter space.
In this study,  we expand this search by adopting the proton synchrotron model for the 
$\gamma$-ray blazar emission. 

{ Under the assumption that the high-energy component of the SED in FSRQs
is  explained in terms of proton synchrotron radiation,
we showed that for low enough values of the Doppler factor $\delta$, 
the emission from the EM cascade may exceed the observed $\sim$GeV flux. In fact, the 
superposition of the EM cascade and proton synchrotron components results in a spectral hardening
of the total emission above a few GeV (see Fig.~\ref{fit_noext}). 
From our analysis, it 
is not clear  why the proton synchrotron component should dominate in the hard X-ray/soft $\gamma$-ray regime 
while the EM cascade should be suppressed. One could then naturally pose the following question: 
{\sl Would it be possible to fit the SED with the cascade component instead of the proton synchrotron component, 
and would the parameters of such a model be reasonable?} 
Roughly speaking,
the peak of the cascade spectrum can be found by the condition $\tgg(\epsilon_\star)\approx 1$ (for more details, see 
\cite{mannheim93}). For the parameter values used throughout the text,
we showed that $\epsilon_\star \sim 18$~GeV (see eq.~\ref{estar}). In principle,
one could find parameter values that could bring $\epsilon_\star$
down to a few MeV, since $\epsilon_{\star}\propto L_{\rm s}^{-2}R^2 \delta^8 \es$.
In this scenario, the peak luminosity of the cascade should be higher in order to explain 
the observed peak $\gamma$-ray luminosity, namely
 $L_{\pi^0\rightarrow 2\gamma}^{\rm abs}\simeq \tau_{\pg}L_{\rm p} \simeq L_{\gamma}$.
Unless $\tau_{\pg}\gtrsim 1$, which would correspond to 
$R\sim 10^{14}$~cm, $B\lesssim  1$~G, and $\delta \lesssim 5$ (see eq.~(\ref{tpg3})), this scenario
would require higher proton luminosities, and therefore, even more extreme jet powers than those
listed in Table~\ref{tab-2}. 
These estimates, however, are made
by considering only the internal synchrotron photons as targets for both photopion interactions and $\gamma \gamma$
absorption. A proper answer to the question posed above requires 
a self-consistent calculation of the 
cascade emission by taking into account 
the external photon fields 
as targets for both $\gamma \gamma$ and photopion interactions and 
by focusing on a different parameter regime than the one considered here. Such an investigation
is interesting on its own, and will be the subject of a future study.}

{
A serious challenge to the proton synchrotron model arises 
from the need for high
values of $B$ at distances far from the central black hole. 
\cite{Pushkarev2012} have determined, using core shift measurements,
the magnetic field - distance relation to be $B \sim 0.4~\mathrm{G} 
\times (1 \mathrm{pc} / r)$ \citep[see also][for a theoretical investigation]{zdziarskietal14}.
 \cite{Savolainen2008} have presented specifically for 3C~273 
magnetic field measurements of its pc-scale inner jet structure.
Assuming that the jet's angle to our line of sight is
$10^{\circ}$ \citep{Stawarz2004}, it would appear that $B \leq 8$~G  
at $r \simeq 1.25$~pc. Applying the linear relation between $B$
and $r$ and using $\rblr = 0.3$~pc, we find that for the values of
$B$ considered in Fig.~\ref{xmin} and Fig.~\ref{xmin2} our derived
values of $r/ \rblr$ are over a factor of 5 higher than those needed
to accommodate such strong magnetic fields. This would imply that,
at least for the case of 3C~273, 
the magnetic field strength required by the proton synchrotron model
at the location of the emission region is in conflict
with the observations.}

{
In principle, the parameter
space used in modelling the SED could be further constrained by requiring that the jet power ($\pjet$) should not exceed
the accretion power ($\pacc$) which, for FSRQs like 3C~273, can be safely estimated. We showed, however,
that in the proton synchrotron model for 3C~273 the jet power exceeds that of accretion, i.e. 
$\pjet \gtrsim 10\pacc$, for all reasonable parameter values. 
We note, however, that this is as much a problem for leptonic 
models as for hadronic ones \citep{ghisellini14}. }

A possible caveat of our analysis is our assumption of a 
constant radius for the emission region, as well as of a constant
magnetic field strength. Assuming that the emission region
fills the entire cross section of the jet, one could express the size R as a function of distance $r$ from
the central black hole, i.e. $R \propto r^{s}$ with $s>0$ depending on the specific model
of the jet structure \citep[e.g.][]{ghisellini85, moderskietal03,potter13}.  
Similarly, the magnetic field could be written as $B\propto r^{q}$ with $q<0$ \citep[e.g.][]{vlahakis04, komissarov07}.
Yet, there would remain some arbitrariness
regarding the relation between the magnetic field in the jet and in the emission
region, as this would depend on the dissipation mechanism \citep[for a discussion, see][]{sironipetro15},
which in turn depends on the distance from the super-massive black hole \citep[e.g.][]{sikoraetal05, gianniosetal09, nalewajko12}.
In any case, we can qualitatively predict the effects of an increasing
size and decreasing magnetic field strength on the results presented so far.
On the one hand, a larger $R$ would loosen the EM cascade constraint, while
it would push $\delta_{\rm var}$ to higher values. This could bring the location
of the emission region closer to or inside the BLR, at the cost
of even  higher jet powers. On the other hand, a decreasing magnetic field
would make the EM cascade more stringent, i.e. higher $\delta_{\min}$ would
push the location of the emission region even further {away} from the BLR. The impact on jet
power would {depend} on the Doppler factor but it would be marginal, as evidenced by eq.~(\ref{pjet}).
In {short}, an increasing radius would have the opposite effects from those
of a decaying magnetic field. Thus, their combined effect would strongly depend
on the details of the model, such as $q$ and $s$.

The question naturally arises whether {the method presented here} would 
yield significantly different results when
applied to other FSRQs, or to a flaring period of 3C~273.
Our results regarding the high jet power
are not expected to differ, since this is an intrinsic feature
of the proton synchrotron model \citep[e.g.][]{boettcherreimer13}. 
It is not straightforward, however, what 
the answer would be regarding the location of the emission region
in other luminous blazars. The hypothesis
that the emission region, in the hadronic framework for FSRQs, is 
located at the $\sim$pc scale jet, can be easily tested
by applying our method to a sample of densely monitored sources, and we plan to do so 
in a future study.
%

\section{Conclusions}
\label{conclusions}
We  have demonstrated a method of constraining the properties of the $\gamma$-ray emitting
region in FSRQs within the one-zone proton synchrotron model.  
{Even though the high-energy component of the  
blazar SED is attributed to synchrotron radiation of relativistic protons, 
the emission from photohadronic processes cannot be avoided. In fact, 
the EM cascade initiated by the absorption of photons produced
via photohadronic interactions
 may exceed the observed $\gamma$-ray flux at $\sim$GeV energies, for small enough  Doppler factors.
For the purposes of our analytical treatment, we focused on the photons from neutral pion decay, while
we neglected the synchrotron emission from Bethe-Heitler pairs and those produced through the decay of 
charged pions. Therefore, the EM cascade results only from photons from neutral pion decay, which
interact with the low-energy synchrotron blazar emission.
To avoid a significant
alteration of the SED in that energy range, a constraint is set
on the luminosity from the EM cascade which is translated to 
a lower limit on the Doppler factor given by eq.~(\ref{dmin1}).
Our analytical calculations were performed in the 
asymptotic limit where the external radiation from the broad line region (BLR) is negligible. In this regard,
$\dmin$ can be considered as an absolute lower limit.}

{We then applied our method to a single FSRQ, namely 3C~273, 
and generalized our analytical method by including 
the radiation from the BLR, as well as the variability timescale
($\tvar$) in our calculations.  Because 
the BLR energy density in the comoving frame of the emission region 
is highly dependent on the Doppler factor and its 
distance from the super-massive black hole
in the galactic center, we used the EM cascade argument to determine 
a minimum distance for each Doppler factor value. Finally, 
for a given source size, $\tvar$ sets an additional lower limit on the Doppler
factor ($\delta_{\rm var}$).}
With those additional elements, we
arrived at the following robust results: (i) the  Doppler factor of the emission region 
should be higher than $18-20$ for magnetic field strengths $\lesssim 30$~G
and $\sim$day timescale variability; (ii) the $\gamma$-ray emission region should
be located outside the BLR, namely at $r \gtrsim 10\rblr \sim 3$~pc; (iii)
shorter variability timescales, e.g. $\lesssim 12$~hr, push
both the minimum Doppler factor and distance to even higher values; { (iv) the magnetic
field strength required by the model at pc~scale distances is stronger than that
inferred from observations;} and (v)
the jet power exceeds by at least one order of magnitude the FSRQ accretion power.
{In conclusion, our results disfavour 
the proton synchrotron model for the FSRQ 3C 273.}
}

\section*{Acknowledgments}
We thank {the anonymous referee for the insightful comments that helped us clarify
subtleties in the manuscript.} We also thank 
Prof.~D.~Giannios for fruitful discussions and Prof.~A. Mastichiadis 
for comments on the manuscript.
MP acknowledges support by NASA 
through Einstein Postdoctoral 
Fellowship grant number PF3~140113 awarded by the Chandra X-ray 
Center, which is operated by the Smithsonian Astrophysical Observatory
for NASA under contract NAS8-03060.
 We acknowledge the use of data 
from the ASI Science Data Center (ASDC), managed
by the Italian Space Agency (ASI). 
\bibliographystyle{mn2e} 
\bibliography{3c273}


\end{document}